\documentclass[review]{elsarticle}

\usepackage{lineno,hyperref}
\modulolinenumbers[1]

\journal{Journal of \LaTeX\ Templates}
\usepackage{amsmath}

\usepackage{color}
\usepackage{booktabs}
\usepackage{makecell}

\begin{document}

\begin{frontmatter}

\title{Optimization of Timepix3-based conventional Compton camera using electron track algorithm}
\author[PRI,DEP,PPL]{Jiaxing Wen}

\author[PRI,DEP]{Xutao Zheng}

\author[PRI,DEP]{Huaizhong Gao}

\author[PRI,DEP]{Ming Zeng\corref{correspondingauthor}}
\cortext[correspondingauthor]{Corresponding author}
\ead{zengming@tsinghua.edu.cn}

\author[DEP]{Yuge Zhang}

\author[PPL]{Minghai Yu}

\author[PPL]{Yuchi Wu}

\author[DA,DEP]{Jirong Cang}

\author[DEP]{Ge Ma}

\author[PPL]{Zongqing Zhao}

\address[PRI]{Key Laboratory of Particle and Radiation Imaging (Tsinghua University), Ministry of Education, Beijing 100084, China}
\address[DEP]{Department of Engineering Physics, Tsinghua University, Beijing 100084, China}
\address[PPL]{Science and Technology on Plasma Physics Laboratory, Laser Fusion Research Center, CAEP, Mianyang 621900, Sichuan, China}
\address[DA]{Department of Astronomy, Tsinghua University, Beijing 100084, China}

\begin{abstract}
The hybrid pixel detector Timepix3 allows the measurement of the time and energy deposition of an event simultaneously in each 55 $\mu$m pixel, which makes Timepix3 a promising approach for a compact Compton camera. However, the angular resolution of Compton camera based on this kind of detector with high pixel density is usually degraded in imaging of MeV gamma-ray sources, because the diffusion of energetic Compton electron or photoelectron could trigger many pixels and lead to an inaccurate measurement of interaction position. In this study, an electron track algorithm is used to reconstruct the electron track and determine the interaction point. \textcolor{black}{An demonstrative experiment was carried out, showing that the effect of this algorithm was significant. The angular resolution measures of a single layer Compton camera based on Timepix3 was enhanced to $12$ degrees (FWHM) in imaging of a $\rm ^{60}Co$ gamma-ray source.}
\end{abstract}


\end{frontmatter}

\section{Introduction}
\label{sec:intro}
Compton camera has been regarded as one of the most promising technologies for gamma-ray imaging and polarimetry\cite{takeda_polarimetric_2010}\cite{mitani_prototype_2004} in the MeV and sub-MeV energy band, thus have been applied in various fields, such as astrophysics\cite{tajima_design_2005}\cite{watanabe_sicdte_2014}, medical imaging\cite{yamaguchi_spatial_2009}\cite{kabuki_development_2007}\cite{vernekohl_feasibility_2016}, and radiation monitoring\cite{jiang_prototype_2016}\cite{watanabe_development_2018}. The angular resolution of Compton camera is significantly affected by the measurement accuracy of scattering position, absorption position, energy of Compton electron, and energy of scattered photon\cite{ordonez_angular_1999}\cite{ordonez_dependence_1997}. Therefore, employing pixelated semiconductor detectors, with capacities of high energy resolution and spatial resolution, can dramatically improve the angular resolution and hence the sensitivity. \textcolor{black}{Thus,} pixelated semiconductor detectors have been extensively adopted in Compton camera systems recently \cite{watanabe_sicdte_2005}\cite{du_evaluation_2001}\cite{vetter_high-sensitivity_2007}. Timepix3\cite{poikela_Timepix3_2014}, as the state-of-the-art hybrid semiconductor detector readout chip, is considered a good device for building Compton cameras. In recent years, Timepix3-based Compton cameras have been proposed and demonstrated \cite{turecek_compton_2018}, and the functionality of a single layer Timepix3 Compton camera has also been verified\cite{turecek_single_2020}\cite{amoyal_development_2021}.

Timepix3 has a pixel matrix made of 256$\times$256 square-shaped pixels with 55 $\mu$m pitch and can be hybridized to sensors from different semiconductor materials such as Si, CdTe, or GaAs with thicknesses up to a few millimeters. Timepix3 offers good energy resolution of $\sigma$=4 keV at 59.5 keV as well as high time resolution of 1.56 ns\cite{frojdh_Timepix3_2015}. \textcolor{black}{Moreover,} Timepix3 can measure the charge via Time-over-Threshold (ToT)\cite{pitters_time_2019} and the depth of interaction position via time-of-arrival (ToA) in each pixel simultaneously. Timepix3 can be used similarly to a time projection chamber (TPC). The position in the pixel plane of the event can be obtained directly from the location of the triggered pixel, and the position out of the pixel plane (depth) can be determined by ToA. Therefore, Timepix3 can provide the possibility of the three-dimensional (3D) particle trajectory reconstruction or 3D interaction point reconstruction with high spatial resolution\cite{bergmann_3d_2017}\cite{bergmann_3d_2019}\cite{turecek_usb_2016}.


A 3D position-sensitive detector generally requires a small pixel pitch or high spatial resolution. In finely segmented detectors, the energy deposition by the electron is not confined to a detector element, i.e., the electron can trigger many pixels and leave a track in the detector, and the \textcolor{black}{determination} of the initial direction of an electron is possible. With the knowledge of the initial direction of Compton electron, the direction of the incident photon can be reconstructed completely for a single photon, and this kind of Compton camera is called the advanced Compton camera or electron-tracking Compton camera (ETCC)\cite{takada_development_2005}\cite{plimley_reconstruction_2011}\cite{shimazoe_electron_2016}. However, the 55 $\mu$m pixel pitch of Timepix3 is still not sufficient to provide a reasonable resolution for the initial direction of the sub-MeV electron in semiconductor materials\cite{plimley_angular_2016}. On the other hand, it has been shown in recent progress of the X-ray polarimeter, that the reconstruction accuracy of the photon absorption point can be impressively improved even with two-dimensional (2D) photoelectron tracks\cite{bellazzini_direct_nodate} and achieved impressive results\cite{feng_re-detection_2020}\cite{costa_efficient_2001}\cite{huang_simulation_2021}, e.g., a spatial resolution of 28.1 $\mu$m and a modulation factor of 50.9 $\%$ have been achieved with a 50 $\mu$m pixel pitch for 4.5 keV photons (Ti K photo-peak)\cite{noauthor_measurement_2013}. Therefore, in a conventional Compton camera, based on the pixelated semiconductor detector, electron tracks are still expected to contribute to the reconstruction of \textcolor{black}{the} interaction point and the improvement of \textcolor{black}{the} angular resolution.

In the presented work, an electron track algorithm is adopted to better reconstruct interaction points to improve the quality of the Compton camera images. The pixel calibration, clustering method, track reconstruction, and event selection are described in Section\ref{sec:detector_method}. In Section\ref{sec:experiment}, the experimental results of the single layer Timepix3 Compton camera are discussed. The conclusion is given in Section\ref{sec:conclusion}.

\section{Detector calibration and pixel data processing}
\label{sec:detector_method}

\begin{figure}[!htb]
    \centerline{\includegraphics[width=0.7\textwidth]{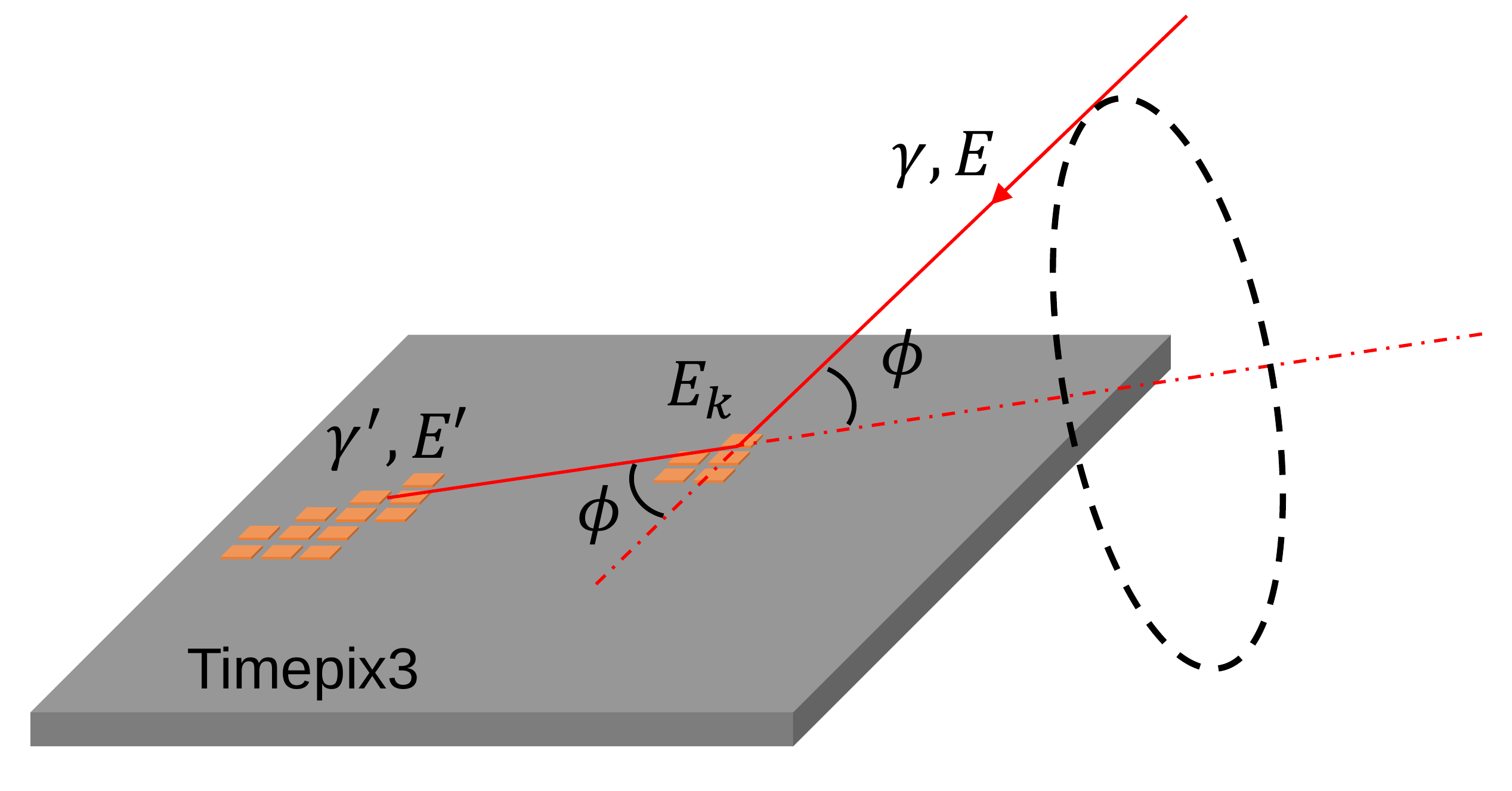}}
    \caption{The schematic of the Timepix3-based single layer Compton camera. An incident photon $\rm \gamma$ with energy $\rm E$ is scattered in the sensor, a Compton photon $\rm \gamma',E'$ and a Compton electron $\rm e,E_{k}$ are detected by the same Timepix3 detector, and the back-projected Compton cone can be determined. One Compton electron and photoelectron can trigger several neighboring pixels in the detector, forming a cluster.}
    \label{fig:ComptonScatter}
\end{figure}

The Timepix3 detector we used is MiniPIX TPX3 manufactured by ADVACAM, which is the same model used in Ref.\cite{turecek_single_2020}, and equipped with a CdTe sensor of 1 mm thickness, 14.08 mm width, and 14.08 mm length. The bias voltage of Timepix3 will significantly affect its performance. With the decrease of the bias voltage absolute value, the charge collection efficiency decreases, leading to a degradation of energy resolution, and the drift velocity of charge carrier decreases leading to a better depth resolution\cite{bergmann_3d_2019}. Therefore, the Timepix3 detector is operated in a bias voltage of -100 V which is a trade-off between the depth resolution and energy resolution. The Timepix3 detector can be used as a single layer Compton camera as shown in Fig.\ref{fig:ComptonScatter}. The Compton electron energy $\rm E_{k}$ and Compton photon energy $\rm E'$ are energies of the two clusters of pixels respectively, and the interaction points can be reconstructed from the pixel data. Since the measurement of interaction time is crucial to determine the time-coincidence Compton events and the interaction depth, the detailed calibration methods and results of pixel interaction time are described in this section. The calibrations of time-walk and drift time to depth are following the methods provided in Ref.\cite{bergmann_3d_2017}\cite{bergmann_3d_2019}\cite{turecek_usb_2016}\cite{turecek_single_2020}. The charge induction time is calibrated experimentally using many muon black(preferably more than 100) tracks in this manuscript, while it is calibrated by a theoretical model in the previous study\cite{bergmann_3d_2019}. Besides, the methods of pixel clustering, interaction point reconstruction, and Compton event selection are also described in this section.

\subsection{Detector calibration}
\label{sec:timewalk}
The reconstruction of depth from ToA is based on the modeling of charge carrier propagation and signal induction in the sensor and Timepix3 readout chip, which is described in detail in the previous studies\cite{bergmann_3d_2019}\cite{filipenko_3d_2014}\cite{bergmann_3d_2017}. The time from interaction to trigger in each pixel, denoted by trigger time $\rm t_{trigger}$, can be expressed as
\begin{equation}
    \rm t_{trigger} =\rm  t_{timestamp}-t_{interaction}\\
    \rm = t_{drift}+t_{time-walk}-t_{induction}.
\label{eq:trigger_time}
\end{equation}
The crucial points of the depth reconstruction are the modeling of the drift time $\rm t_{drift}$ as a function of the interaction depth z, the contribution of time-walk effect $\rm t_{time-walk}$ and the contribution of induction process $\rm t_{induction}$.

The so-called time-walk effect is caused by the different slopes of signals with different amplitudes. Time-walk can be calibrated experimentally\cite{bergmann_3d_2017}\cite{bergmann_3d_2019}\cite{turecek_usb_2016} or by test pulses\cite{frojdh_Timepix3_2015}, and the experimental calibration of time-walk usually uses an $\rm ^{241}Am$ source. The relation between the time delay caused by time-walk and the pixel energy is usually given as Eq.\ref{eq:time-walk}

\begin{equation}
    \rm{t_{time-walk}(E) = \frac{a}{E-b}+d}
    \label{eq:time-walk}
\end{equation}

\begin{figure}[!htb]
\centering
	\includegraphics[width=0.5\textwidth]{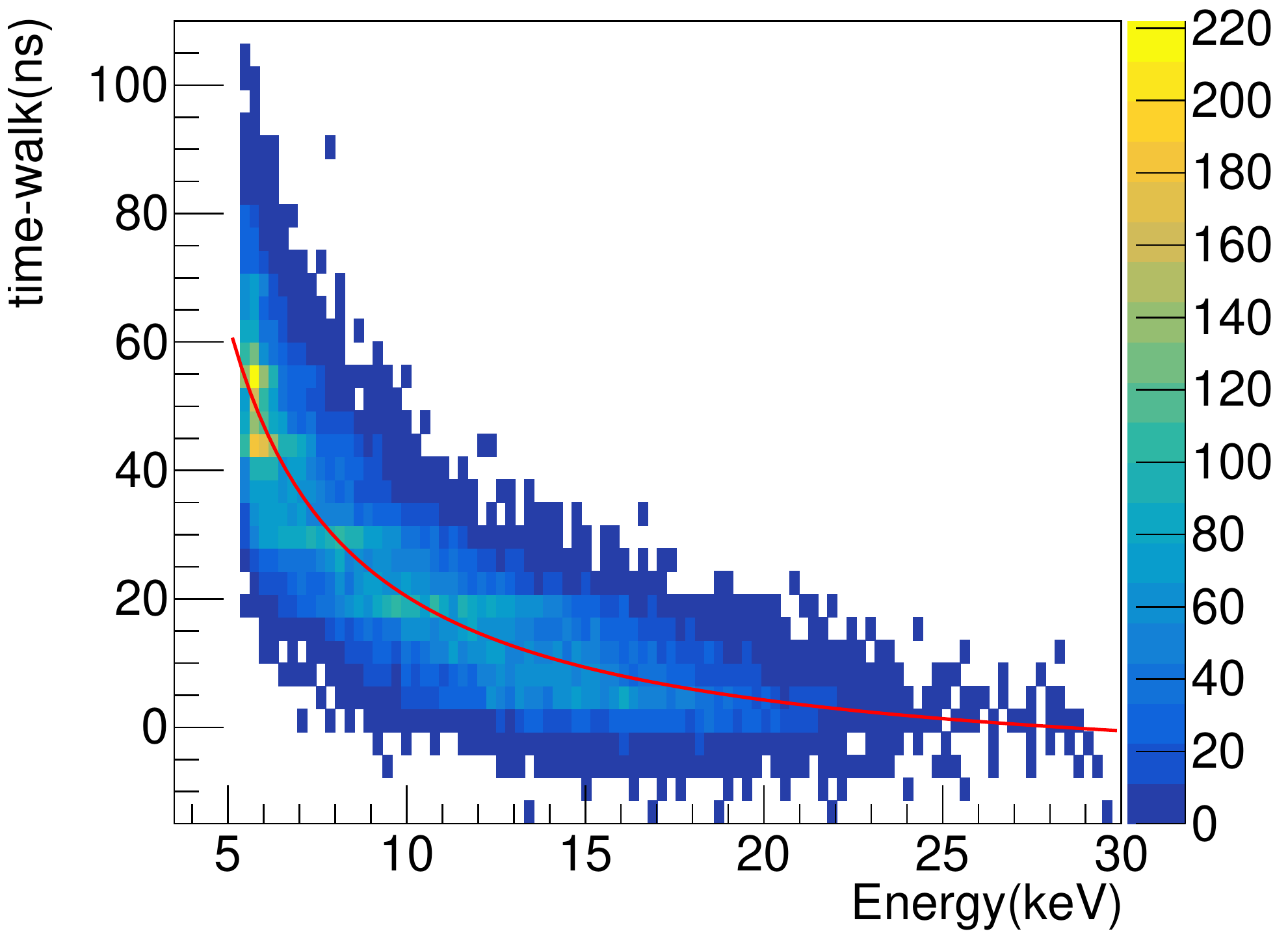}
	\includegraphics[width=0.45\textwidth]{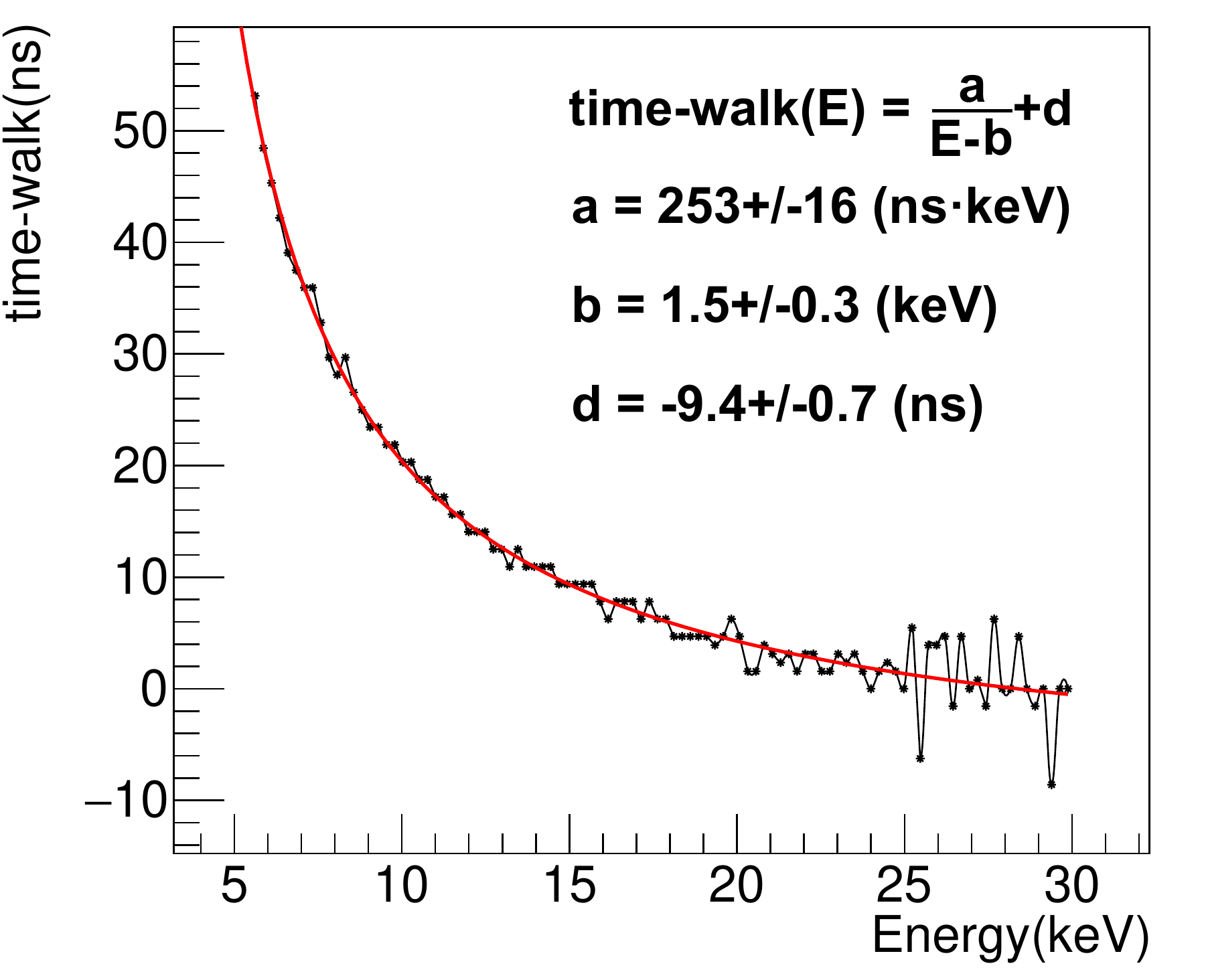}
	\caption{A scatter plot of delays due to the time-walk effect in dependence of the energy of pixels (left). Delay due to time-walk as a function of energy measured in the pixels (right). The function is obtained by fitting the medians of the time delay distributions for each energy channel and plotted as red lines.}
\label{fig:time-walk}
\end{figure}

In this paper, an $\rm ^{241}Am$ source is used for the experimental calibration of time-walk. While the sensor is irradiated with 59.6 keV photons from an $\rm ^{241}Am$ source, 4 pixels can be triggered by one photon if the photon interacts with the sensor near the centre of the 4 pixels. Since the 4 pixels have the same interaction depth, the time delays of trigger time between these pixels are mainly contributed by the time-walk effect. The scatter plot of measured pixel energies $\rm E_i$ versus their time delays $\rm t_{i,time-walk}$ for 7193 4-pixel cluster events are shown in Fig.\ref{fig:time-walk}. We find $\rm a=253\pm16\ ns\times keV$, $\rm b=1.5\pm0.3\ keV$ and $\rm d=-9.4\pm0.7\ ns$ by fitting the medians of the time delay distributions for each energy channel(see Fig.\ref{fig:time-walk}).

\begin{figure}[!htb]
\centering
	\includegraphics[width=0.5\textwidth]{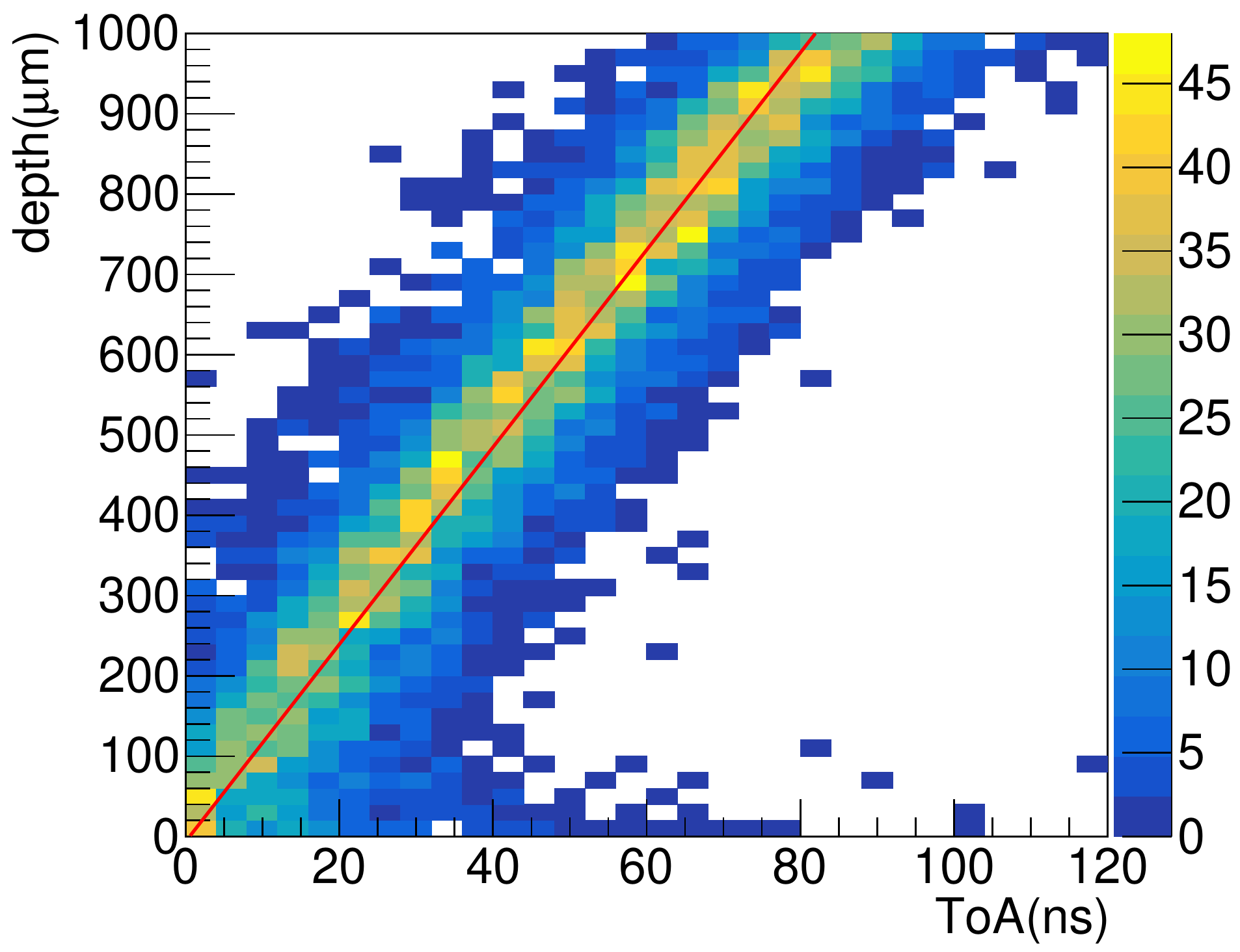}
	\includegraphics[width=0.45\textwidth]{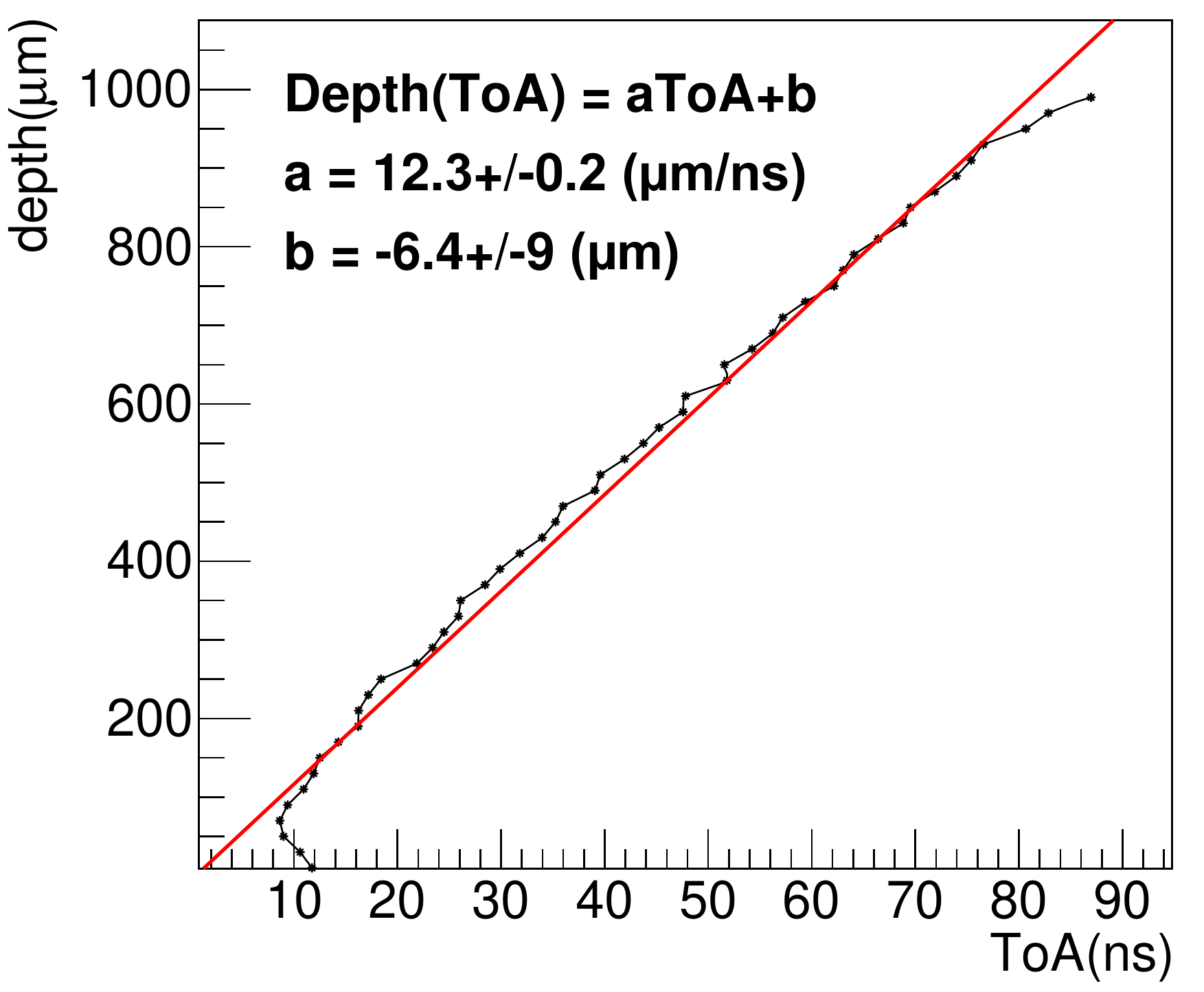}
	\caption{A scatter plot of calculated drift time in dependence of the depth from geometry (left). The medians of drift times distributions for each depth bin as a function of geometry depth (right).}
\label{fig:drift_time}
\end{figure}

A measurement with many cosmic muons is performed to calibrate the drift time to depth. The timestamps of the measured Timepix3 pixel data with energies below 30 keV are corrected for time-walk effect, then, the minimum timestamp in each muon track is regarded as the interaction time. While the time-walk effect is calibrated experimentally, the time delay due to the induction process is automatically compensated for in the energy range below 30 keV\cite{bergmann_3d_2019}. Therefore, the drift time of pixel with energy below 30 keV can be calculated from Eq.\ref{eq:trigger_time}. The cosmic muons have straight paths penetrating the whole thickness of the sensor slantwise, therefore, the interaction depth of each pixel can be calculated from the sensor geometry. In Fig.\ref{fig:drift_time}, the left side shows the scatter plot of the calculated drift time in dependence of the depth from geometry for pixels in 136 muon tracks, and the right side shows the medians of the drift time distributions for each depth bin and the linear fitting of depth and drift time. With this knowledge, the time difference between two pixels can be converted to their depth distance.

The charge induction effect will make the output pulse of the charge-sensitive amplifier reach the trigger threshold before the charge carriers actually arrive at the collecting electrodes. In the previous study\cite{bergmann_3d_2019}, a theoretical model of the charge induction process is used to calculate the time delay caused by the induction process. Here the charge induction time is calibrated experimentally with many cosmic muon tracks. 

\begin{figure}[!htb]
\centering
	\includegraphics[width=0.5\textwidth]{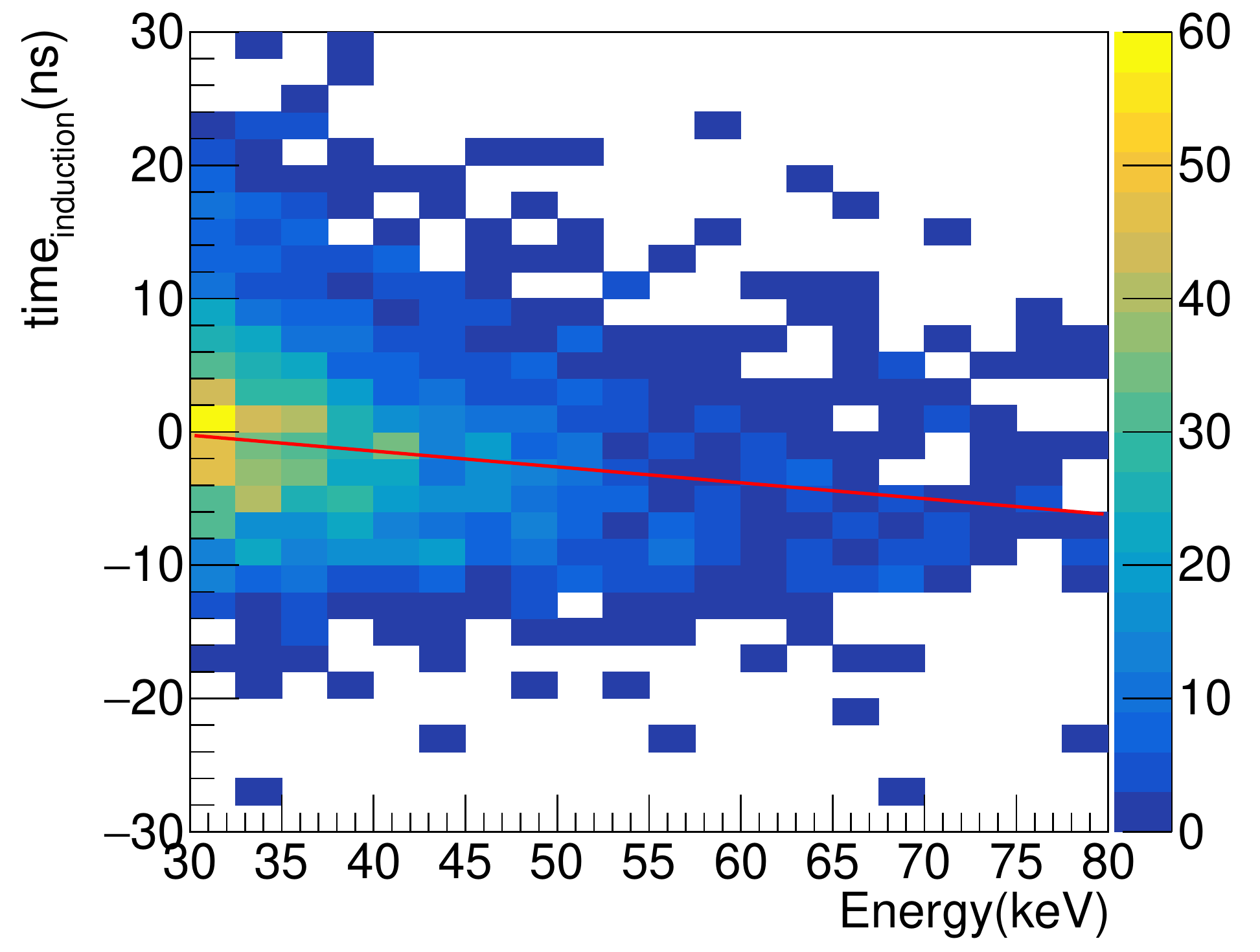}
	\includegraphics[width=0.45\textwidth]{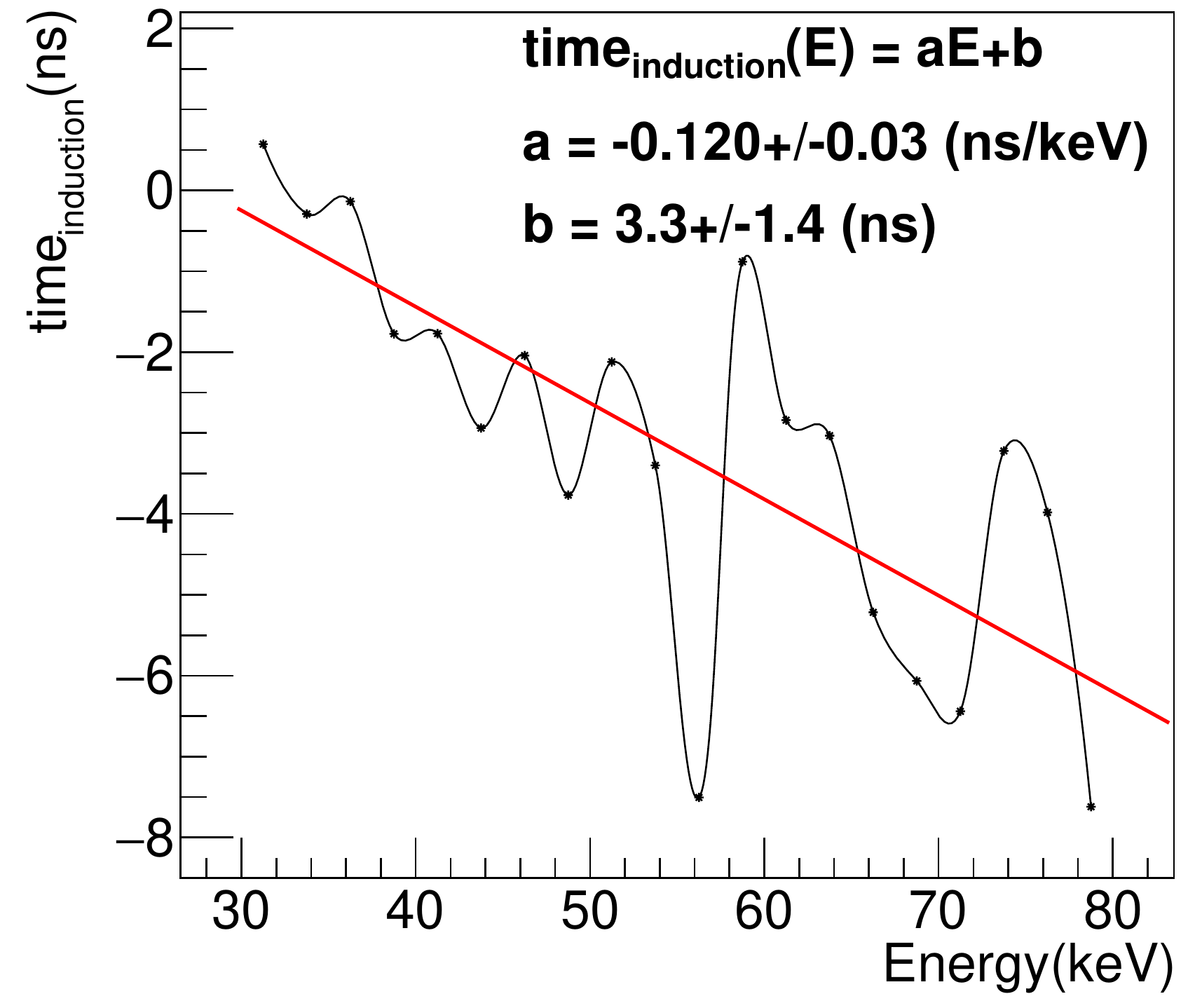}
	\caption{A scatter plot of measured pixel energies $E_i$ versus their time delays due to charge induction (left). The medians of the time delay distributions due to charge induction for each energy channel as a function of energy measured in the pixels (right).}
\label{fig:time_induction}
\end{figure}

In each muon track, the drift time of each pixel can be calculated from its geometry depth according to the depth-drift time relation. In a pixel with energy above 30 keV, the difference between the calculated drift time and trigger time $\rm t_{trigger}-t_{drift}$ is mainly contributed by the charge induction process. Fig.\ref{fig:time_induction} shows the scatter plot of measured pixel energies $E_i$ versus their time delays due to charge induction for 136 muon tracks and the medians of the time delay distributions for each energy channel. A simple linear function is used to fit the induction time and energy, and we find $\rm a=-0.120\pm0.03\ ns/keV$ and $\rm b=3.3\pm1.4\ ns$. The poor fit in the high energy range is due to the small number of events.

The time-walk correction and induction time correction function can be expressed as

\begin{equation}
\begin{split}
&\rm{t_{time-walk}(E)-t_{induction}(E)}= \\
&\left \{
\begin{array}{ll}
    &\rm{\frac{253}{E-1.5}-9.4,\ 5.5\ keV<E<30\ keV}\\
    &\rm{-0.12\times E+3.3,\ E>30\ keV}.
\end{array}
\right.
\end{split}
\label{eq:ref_time-walk}
\end{equation}

\begin{figure}[!htb]
\centering
	\includegraphics[width=0.8\textwidth]{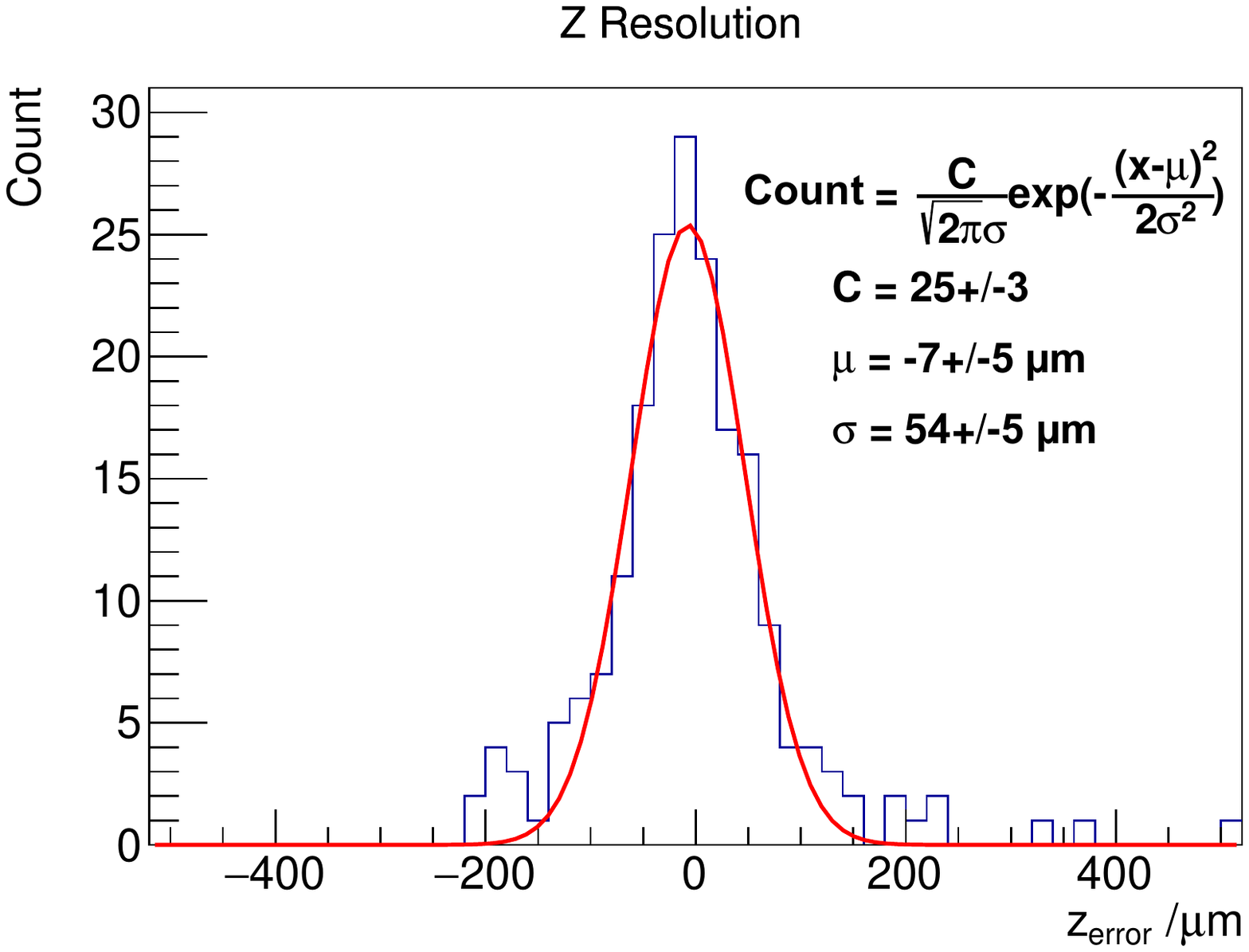}
	\caption{The dispersion of the reconstructed depth around the expected track are described by a gaussian with a width $\rm \sigma_z$ of 54 $\rm \mu m$.}
\label{fig:depth_resolution}
\end{figure}

The depth resolution is determined by comparing the reconstructed depth with the geometry depth in a muon track. The comparison is shown for a muon track with 198 pixels as shown in Fig.\ref{fig:depth_resolution}. The depth resolution is $\rm 54\ \mu m$, while the depth resolution is $\rm 58\ \mu m$ without the induction time correction.

\subsection{Pixel clustering and Compton event selection}
\label{sec:Compton_reconstruction}
Considering that an event may involve many pixels, a clustering algorithm is used to reconstruct the Compton event. The clustering algorithm, Density-Based Spatial Clustering of Applications with Noise (DBSCAN), is adopted in this study, which could find clusters with any shape and discard the low-density area as noise\cite{schubert_dbscan_2017}. \textcolor{black}{There are two parameters in DBSCAN, $eps$ and $min_{samples}$. $eps$ is defined as the maximum generalized distance between two data points for one to be considered as the neighbor of the other, and $min_{samples}$ is defined as the minimum number of neighbors within $eps$ for a point (including the point itself) to be considered as a core point. All the neighboring core points and all of their neighbors will be classified into one cluster. The non-core points, whose all neighbors are not core points, will be regarded as the noise points.}


\begin{figure}[!htb]
\centering
	\includegraphics[width=0.8\textwidth]{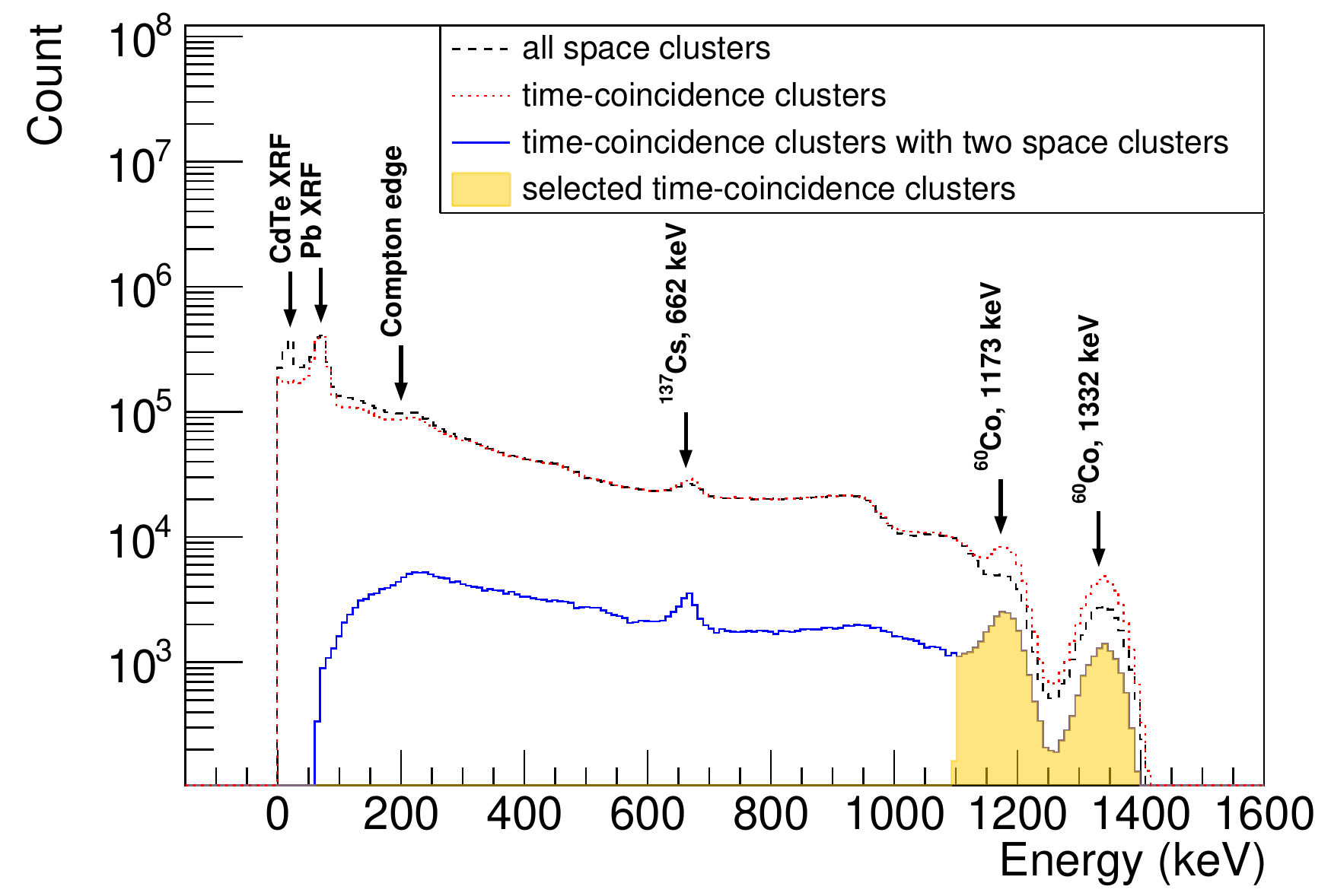}
	\includegraphics[width=0.7\textwidth]{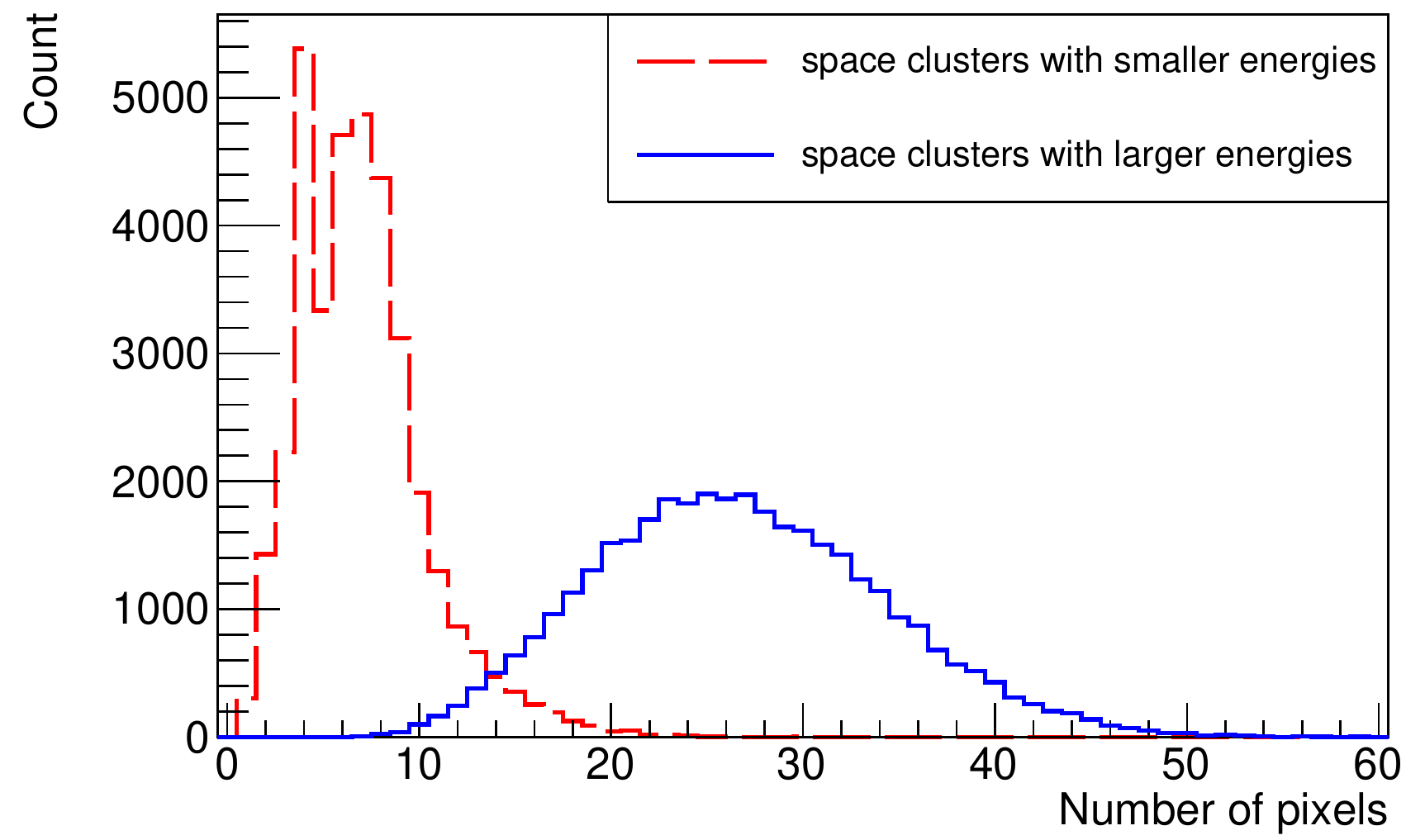}
	\caption{Spectrum of different clusters (top) and distribution of pixel numbers of the selected clusters (bottom).}
\label{fig:event_select}
\end{figure}

There are two steps for the pixel clustering and Compton event selection. First, measured Timepix3 pixel data are divided into many time-coincidence clusters by the DBSCAN algorithm based on the corrected trigger times. Since the $eps$ is the generalized distance between two data points, the $eps$ should be defined as the time interval between measured Timepix3 pixel data in this step. The $eps$ is set to $\rm 1\ \mu s$, which defines the coincidence time window. \textcolor{black}{The $min_{samples}$ is set to one, which means all the pixels are core points and any two pixels with time intervals less than $\rm 1\ \mu s$ will be classified into one time-coincidence cluster.} The pixel with the minimum timestamp in a time-coincidence cluster is considered to have an interaction depth of zero and the interaction depths of other pixels in this time-coincidence cluster can be determined.

Second, pixels in one time-coincidence cluster are grouped into some space clusters by the DBSCAN algorithm based on the 2D spatial distance in the pixel plane. The $eps$ should be the spatial distance between measured Timepix3 pixel data in this step. Therefore, the $eps$ is set to $\rm 156\ \mu m$, which is approximately two times the pixel's diagonal pitch to ensure that different pixels triggered by the same electron are grouped into one space cluster, \textcolor{black}{and the $min_{samples}$ is also set to one in this step}. The space clusters with energies below 30 keV, most of which are signals from internal X-ray fluorescence of the CdTe sensor, are discarded. The time-coincidence clusters with two space clusters are from the internal Compton scattering.

An experiment using point-like $\rm ^{137}Cs$ and $\rm ^{60}Co$ sources was performed for the verification of the pixel clustering method. The spectrum obtained is illustrated on top side of Fig.\ref{fig:event_select}. Compared with the spectrum of all the space clusters, the Compton edge is reduced and the photopeak is improved in the spectrum of the time-coincidence clusters, as shown on the top side of Fig.\ref{fig:event_select}. This indicates that the internal Compton scattering events are correctly detected.


\subsection{Interaction point and track reconstruction}
\label{sec:interaction_reconstruction}
The most important acquisition parameters in a Compton camera are the positions of scattering point and absorption point, the energy deposition of Compton electron, and the total energy \textcolor{black}{deposition} in sensor\cite{maxim_probabilistic_2015}. The energies of the Compton electron and scattered photon can be calculated easily by summing the energies of pixels in the two space clusters, respectively. However, the determination of the interaction point is challenging. The figure on the bottom side of Fig.\ref{fig:event_select} shows the distribution of pixel numbers of the two space clusters in the selected time-coincidence clusters. It can be seen that the Compton electrons and the Compton photons produced by Compton scattering of 1173 keV and 1332 keV gamma-ray photons always trigger many pixels in the Timepix3 detector.

\begin{figure}[!htb]
	\includegraphics[width=0.95\textwidth]{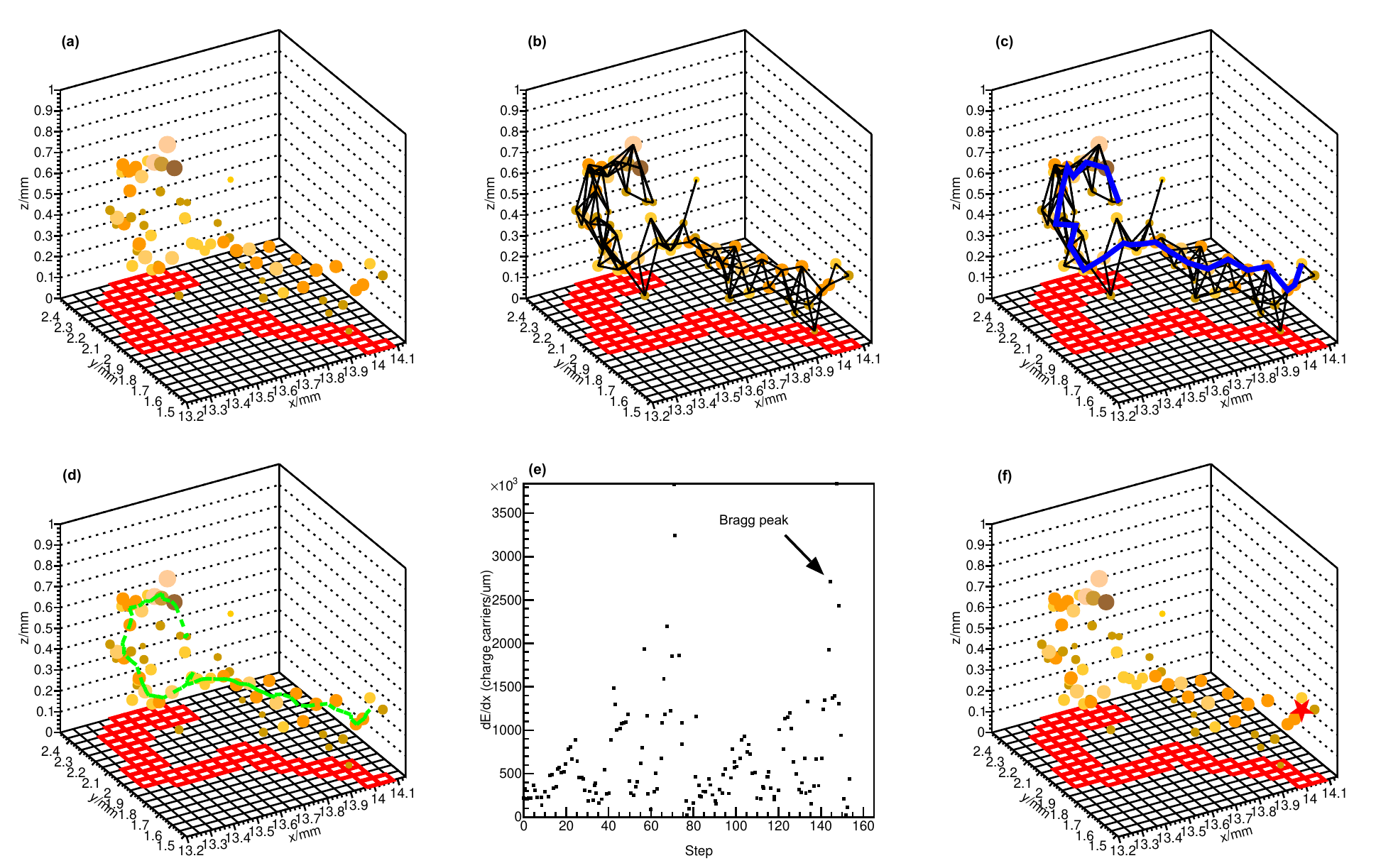}
	\caption{Illustration of the electron track reconstruction algorithm. (a) The reconstructed 3D track of a recoil electron or photoelectron detected by Timepix3. (b) Neighboring points are connected with black solid lines. (c) The primary path (blue line), defined as the longest one of all shortest paths between every two points in the graph. (d) The reconstructed path (green dashed line), derived from the primary path after spatial energy filtering. (e) The reconstructed path is divided into 200 steps, and the energies deposited in each step, dE/dx, are estimated by calculating the sum of the pixel charge around each step. The Bragg peak is clear and the start point (interaction point) and end point of the reconstructed path can be determined. (f) The interaction point reconstructed (red star).}
\label{fig:CSETA2D}
\end{figure}

To reconstruct the interaction point in a long electron track, the electron track algorithm based on the shortest path problem in graph theory, which is described in detail in our previous work\cite{li_electron_2017}\cite{zeng_3-d_2017}, is applied to the space clusters with pixel number above 12. The electron track algorithm reconstructed the electron path and interaction point following the steps shown in Fig.\ref{fig:CSETA2D}. And the charge barycentre method is also used for comparison. In the charge barycentre method, interaction point is determined by calculating the charge barycentre in each space cluster.

\section{Compton imaging experiment}
\label{sec:experiment}
Several measurements were performed with different gamma-ray sources, such as the $\rm ^{137}Cs$, $\rm ^{133}Ba$, and $\rm ^{60}Co$. The activities, characteristic energies used for Compton imaging, and experimental energy resolutions of these gamma-ray sources are listed in Table.\ref{tab:Source}. The goal of these tests is to verify the improvement of angular resolution of the single layer Timepix3 Compton camera when the electronic track algorithm is used to reconstruct the interaction point. Therefore, only a Monte-Carlo back-projection (MC-BP)\cite{yao_study_2020} is used for the reconstruction of the gamma-ray source image.

\begin{table}[htbp]
	\centering
	\caption{The activities, characteristic energies, and measured energy resolutions of the different gamma-ray sources are listed.}
	\begin{tabular}{cccc}
		\toprule
		\makecell[c]{Gamma-ray\\source}&Activity [kBq]&\makecell[c]{Characteristic \\ energies [keV]}&\makecell[c]{Experimental \\ resolutions [$\sigma$ in keV]} \\ 
        \midrule
        $\rm ^{133}Ba$&65.5&356&25.1 \\
        $\rm ^{137}Cs$&8.7&661.5&34.1 \\
        $\rm ^{60}Co$&43.1&\makecell[c]{1173\\1332}&\makecell[c]{53.4\\64.1} \\
		\bottomrule 
	\end{tabular}
	\label{tab:Source}
\end{table}

\subsection{The angular resolution measures}
\label{sec:Co60-onside}
\begin{figure}[!htb]
\centering
	\includegraphics[width=0.8\textwidth]{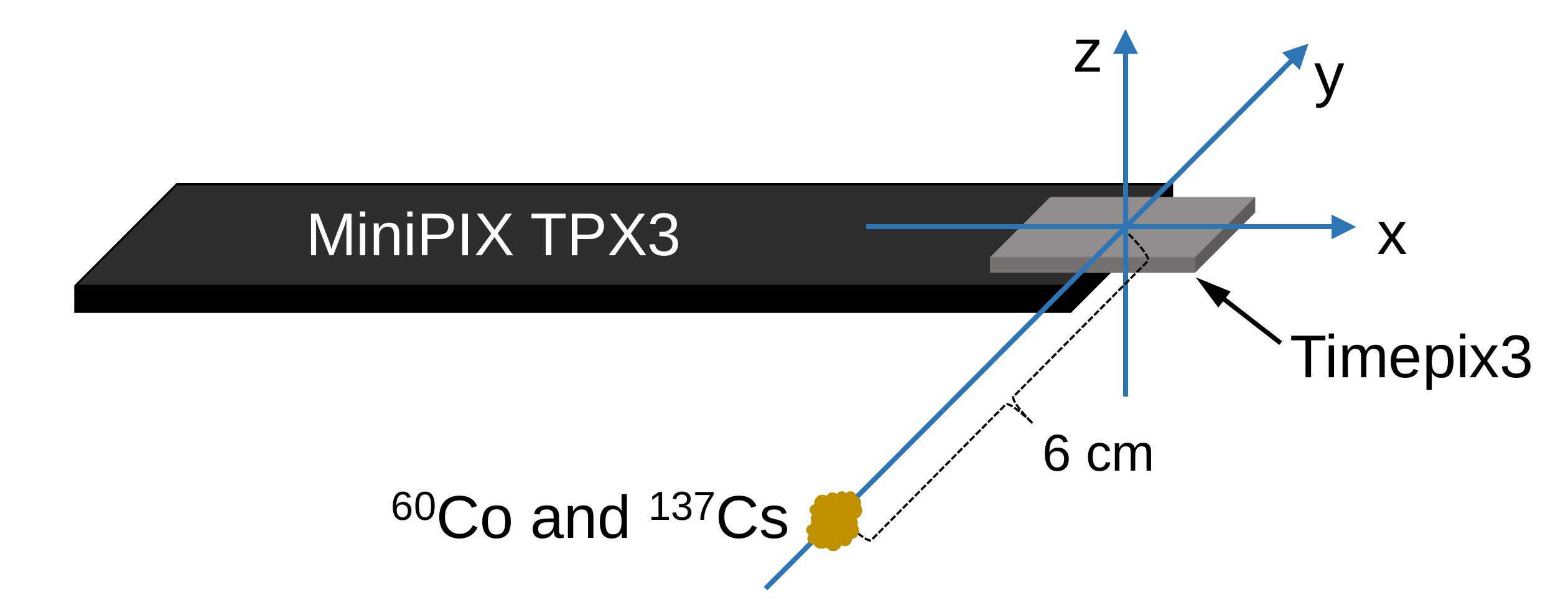}
	\caption{The schema of experimental setup. The sources are placed at the side of the Timepix3 detector in the distance of 6 cm.}
\label{fig:ExperimentTwo}
\end{figure}

The angular resolution measures (ARM) is defined as the difference between the first kinematic scattering angle and the first geometrical scattering angle calculated assuming that the incident direction of the gamma-ray is parallel to the line-of-sight direction\cite{ichinohe_first_2016} and is a common criterion used to evaluate the performance of Compton camera. The first measurement is performed with point-like $\rm ^{137}Cs$ and $\rm ^{60}Co$ sources, and the ARM are calculated. The sources are placed at the side of the Timepix3 detector, i.e., close to the pixel plane, and are approximately 6 cm away from the center of the sensor, as shown in Fig.\ref{fig:ExperimentTwo}. Two energy windows, which are 620-700 keV and 1100-1450 keV, are applied on the measured data. The energy spectrum and the selected events are shown on the top side of Fig.\ref{fig:event_select}. Only the time-coincidence clusters from 1100 to 1450 keV (the events from $\rm ^{60}Co$ source) are further processed because the continuous background contributed by multiple Compton scattering is low in this energy range. 

\begin{figure}[!htb]
\centering
	\includegraphics[width=0.4\textwidth]{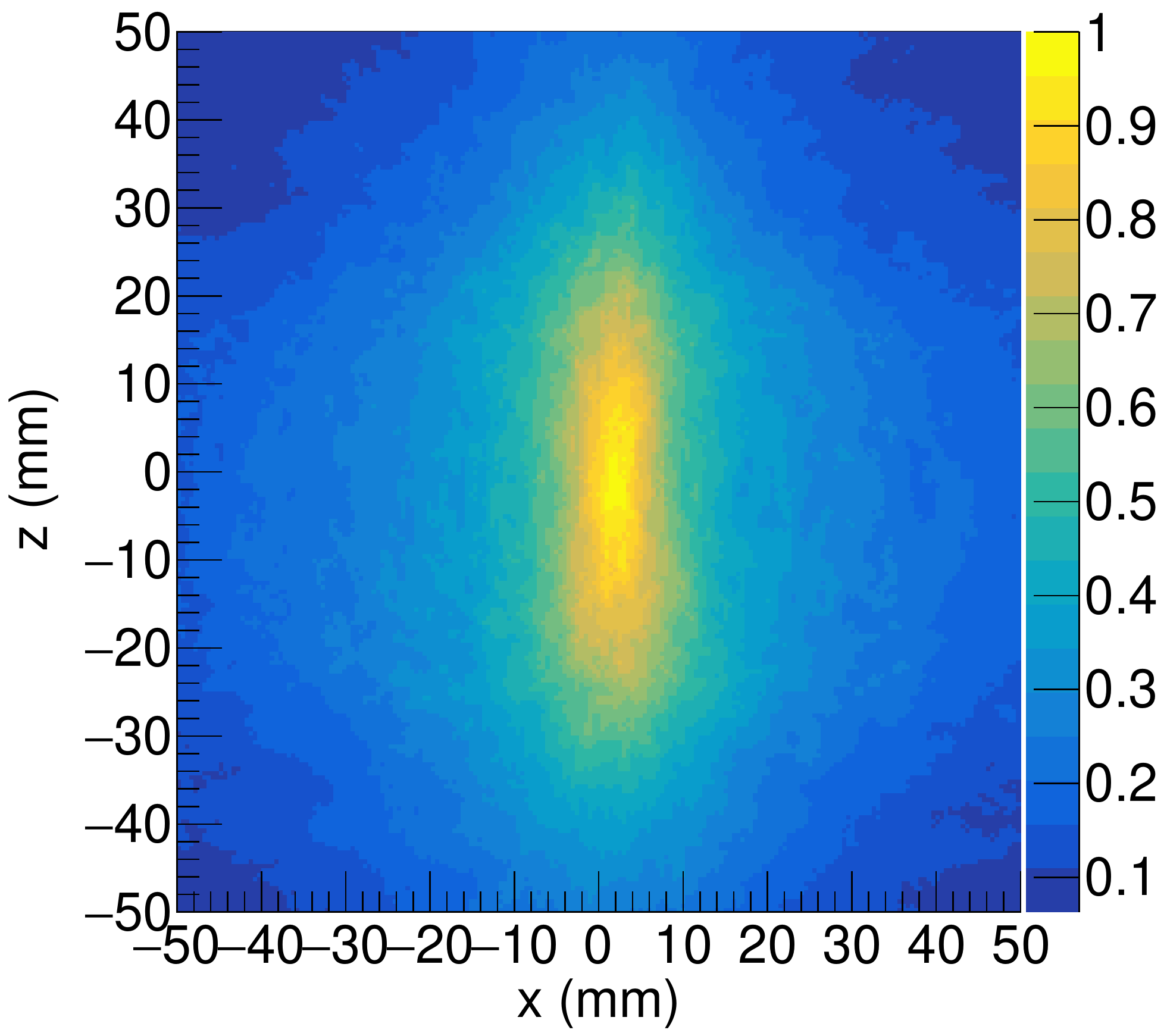}
	\includegraphics[width=0.4\textwidth]{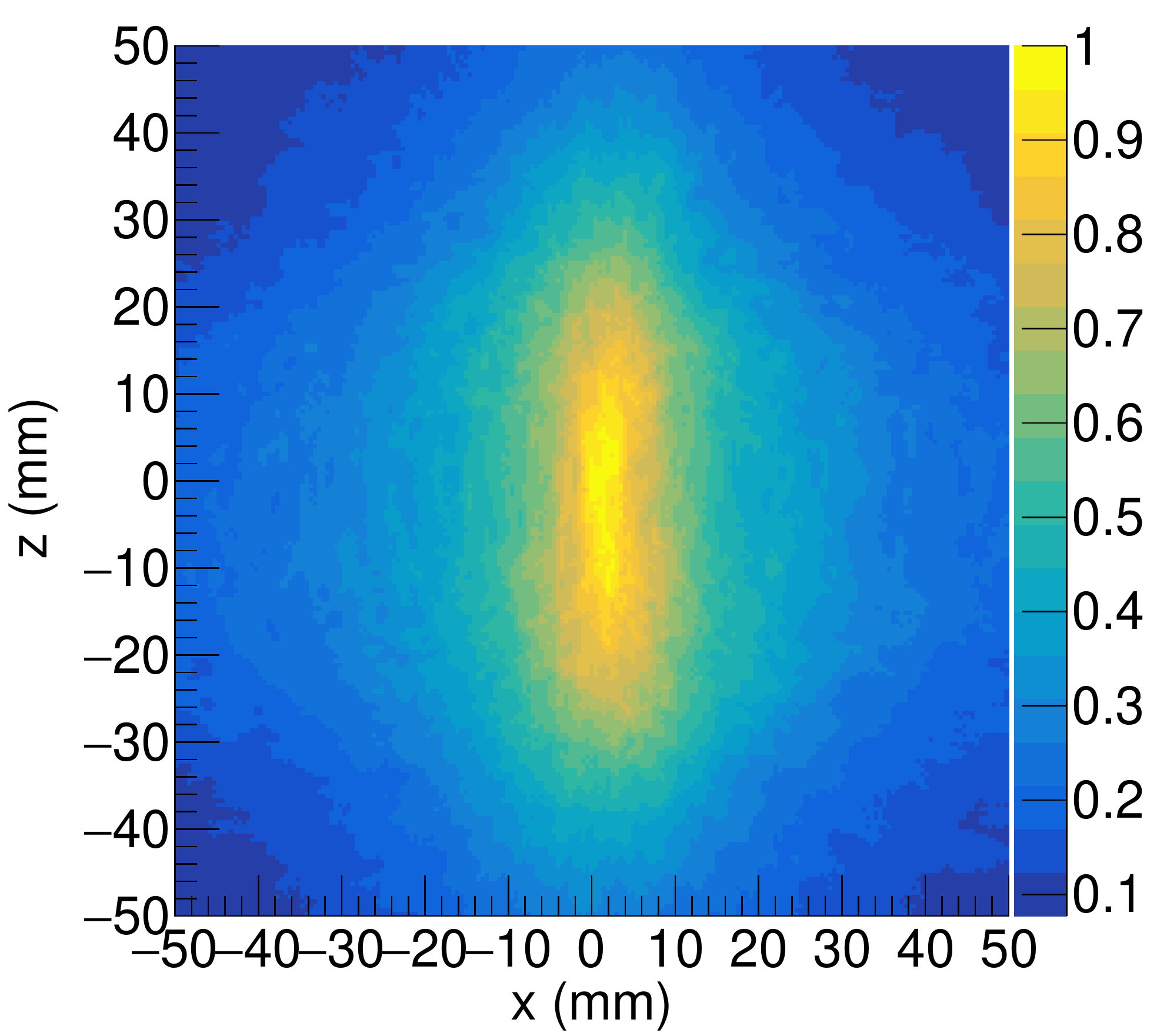}
	\includegraphics[width=0.45\textwidth]{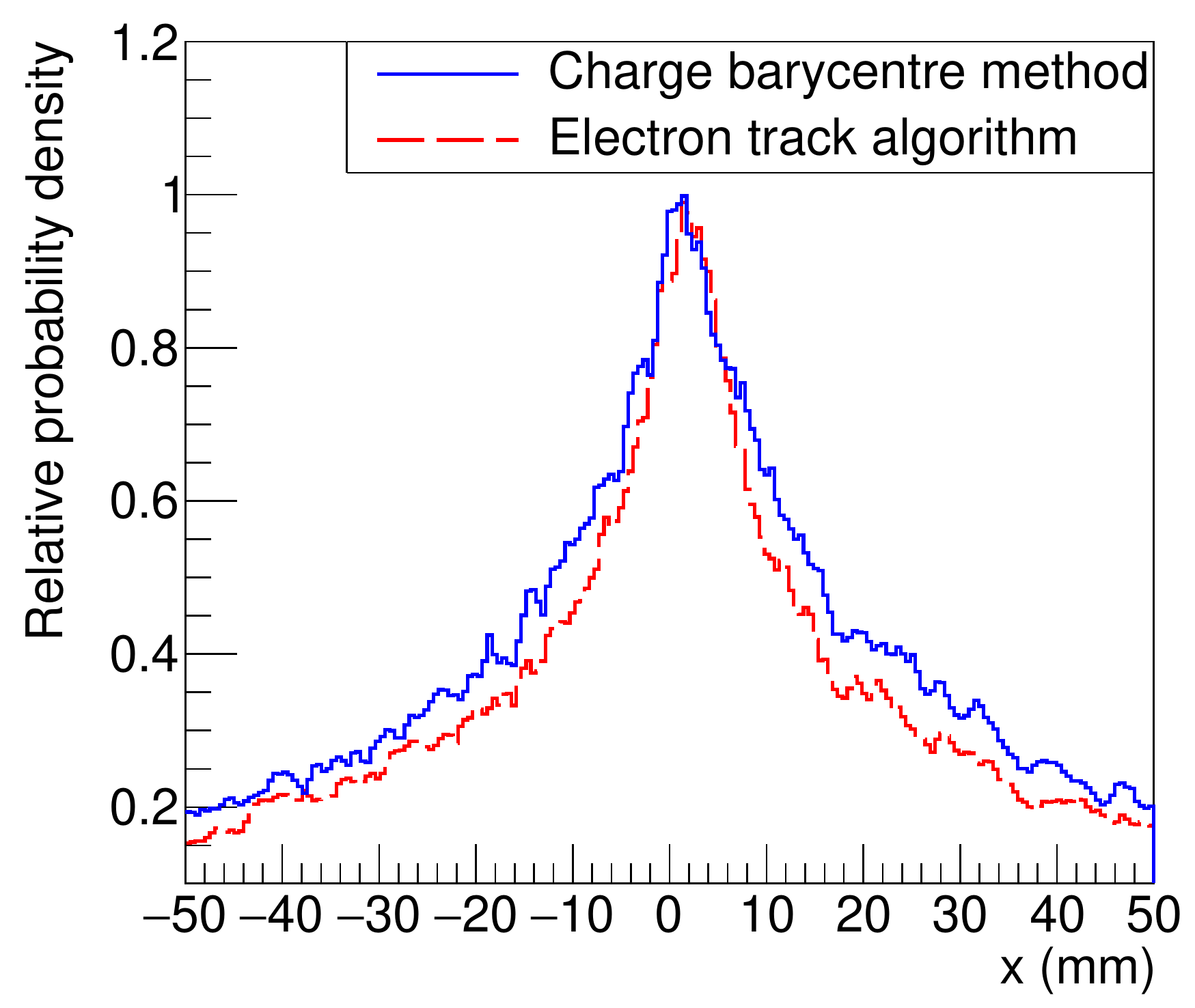}
	\caption{Top left: The interaction points are reconstructed by the electron track algorithm in the space clusters with pixel numbers above 12. Top right: The interaction points are determined by the charge barycentre method in all the space clusters. Bottom: Comparison of one-dimensional (1D) probability density distributions along the x direction with z=0 mm.}
\label{fig:Image_Co60}
\end{figure}

The Compton camera image in a projection plane placed at y=6 cm and parallel to the x-z plane is shown in Fig.\ref{fig:Image_Co60}. The angular resolution of the Compton camera is significantly improved using the electron track algorithm by comparing the one-dimensional profiles of probability density distributions. The full width at half maximum (FWHM) of the distribution has been improved from 28.1 mm to 18.1 mm using the electron track algorithm. The resolution along the z direction is worse than that along the x direction, because most of the scattered photons detected by the single layer detector have the scattering directions parallel to the pixel plane.

\begin{figure}[!htb]
\centering
	\includegraphics[width=0.6\textwidth]{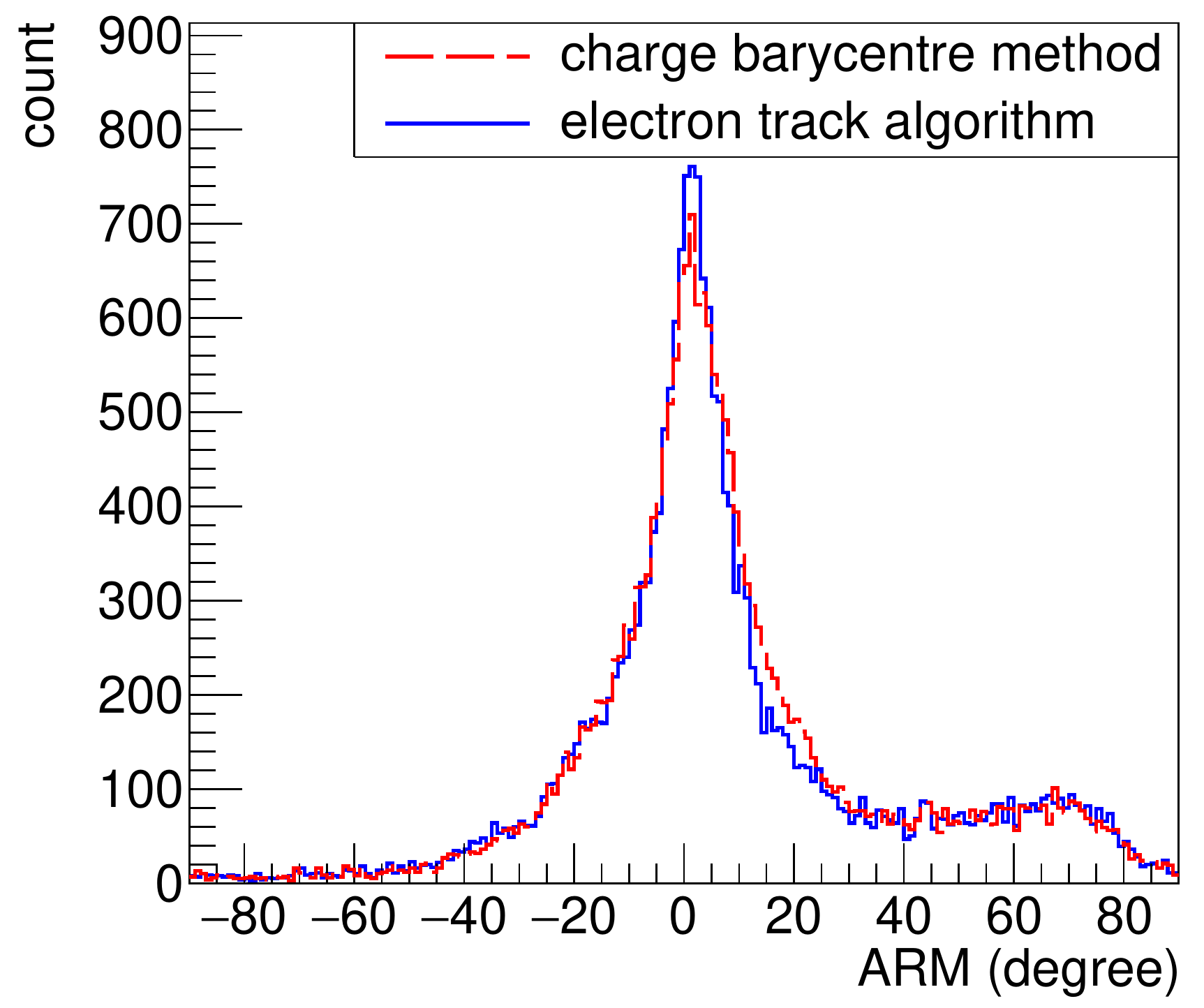}
	\caption{The distribution of ARMs for a $\rm ^{60}Co$ source.}
\label{fig:ARM}
\end{figure}

The distribution of ARMs is shown in Fig.\ref{fig:ARM} which is expected to be a superposition of multiple different distributions involving the contributions of Doppler broadening, energy and spatial resolution\cite{ordonez_angular_1999}\cite{ordonez_dependence_1997}\cite{truemper_doppler_2003}. Therefore FWHM of the ARM distribution is calculated directly from the distribution profile. When the electron track algorithm is applied, the FWHM is 12 degrees. When all the interaction points are determined by the charge barycentre method, the FWHM is 15 degrees. The 20$\%$ improvement of ARM is achieved using the electron track algorithm. The asymmetrical distribution of ARM is caused by the misdetermination of whether space clusters are Compton electron or scattered photon. 

\subsection{Compton camera image of multiple gamma-ray sources}
\label{sec:multi-source}

\begin{figure}[!htb]
\centering
	\includegraphics[width=0.35\textwidth]{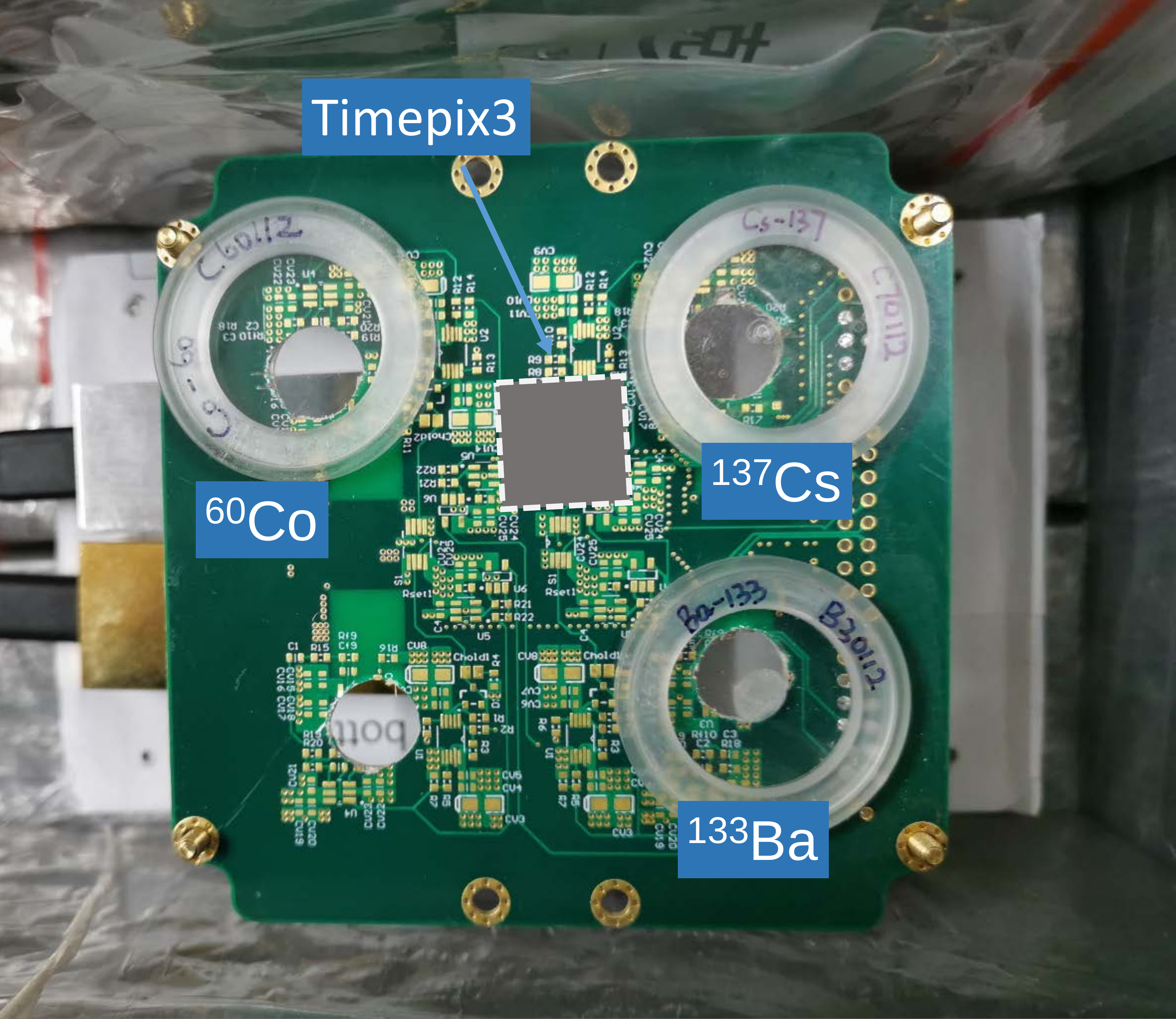}
	\includegraphics[width=0.5\textwidth]{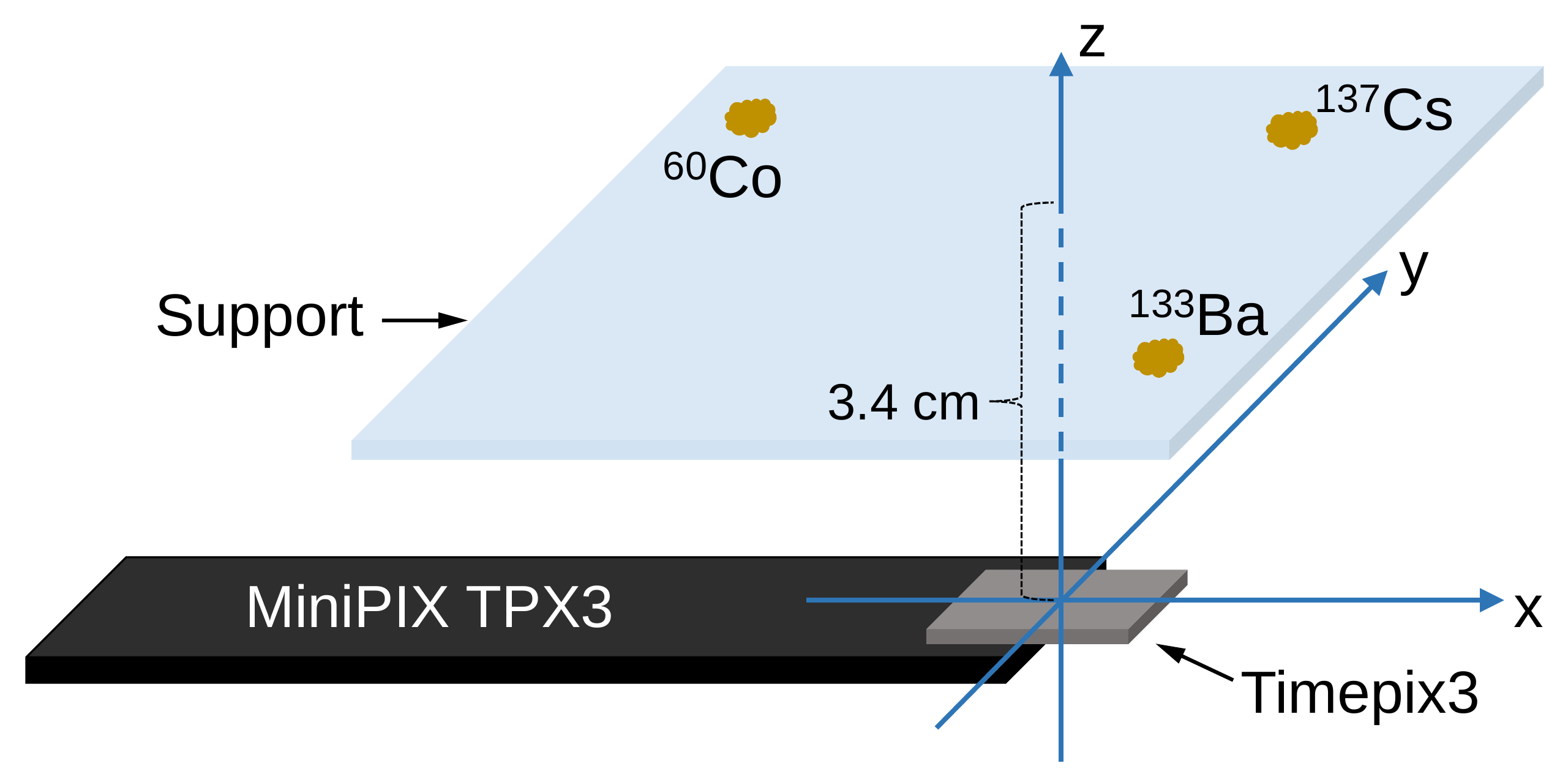}
	\includegraphics[width=0.7\textwidth]{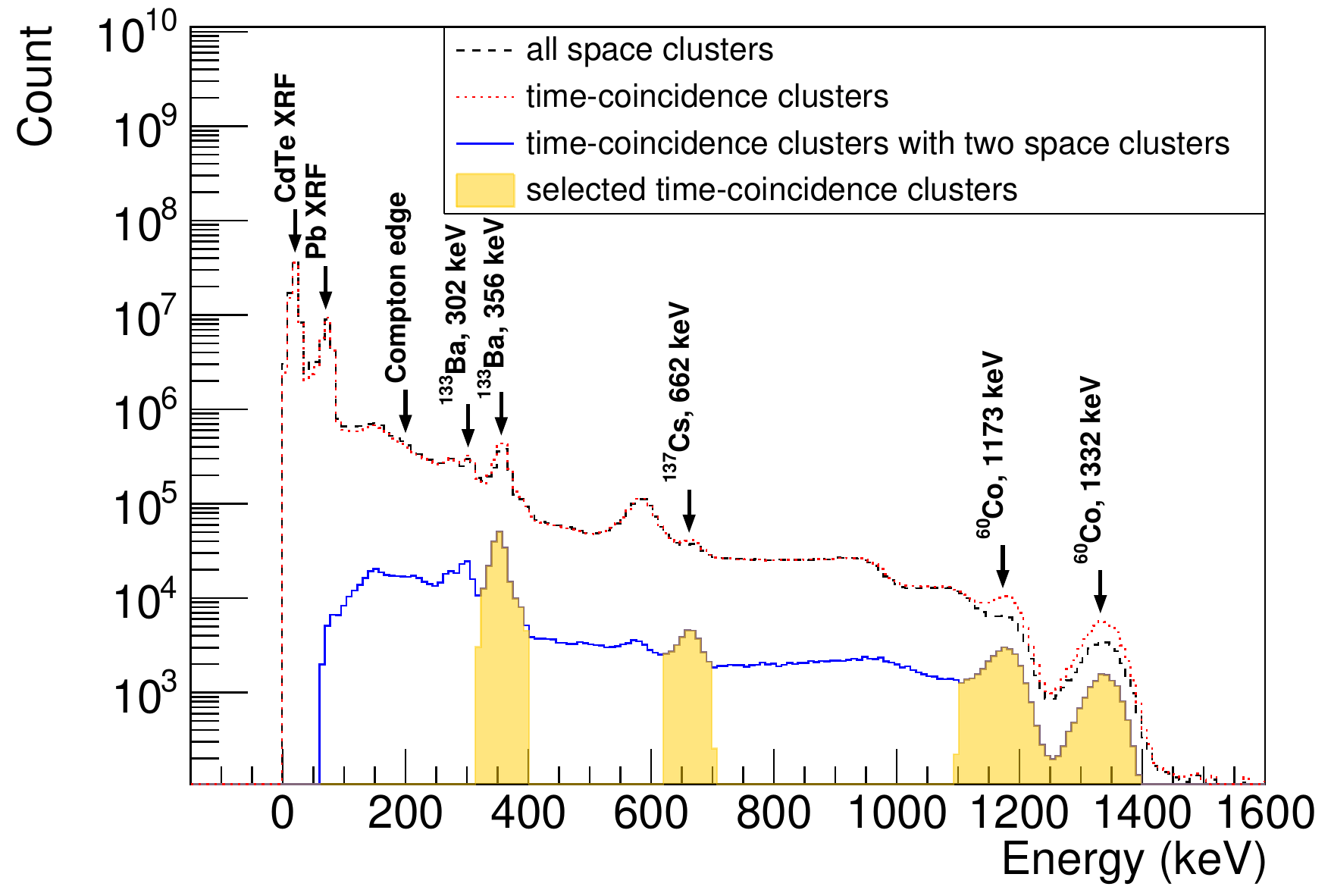}
	\caption{The photography and schema of experimental setup (top). Spectrum of different clusters and the events selected for imaging (bottom).}
\label{fig:ExperimentOne}
\end{figure}

\begin{figure}[!htb]
\centering
	\includegraphics[width=0.3\textwidth]{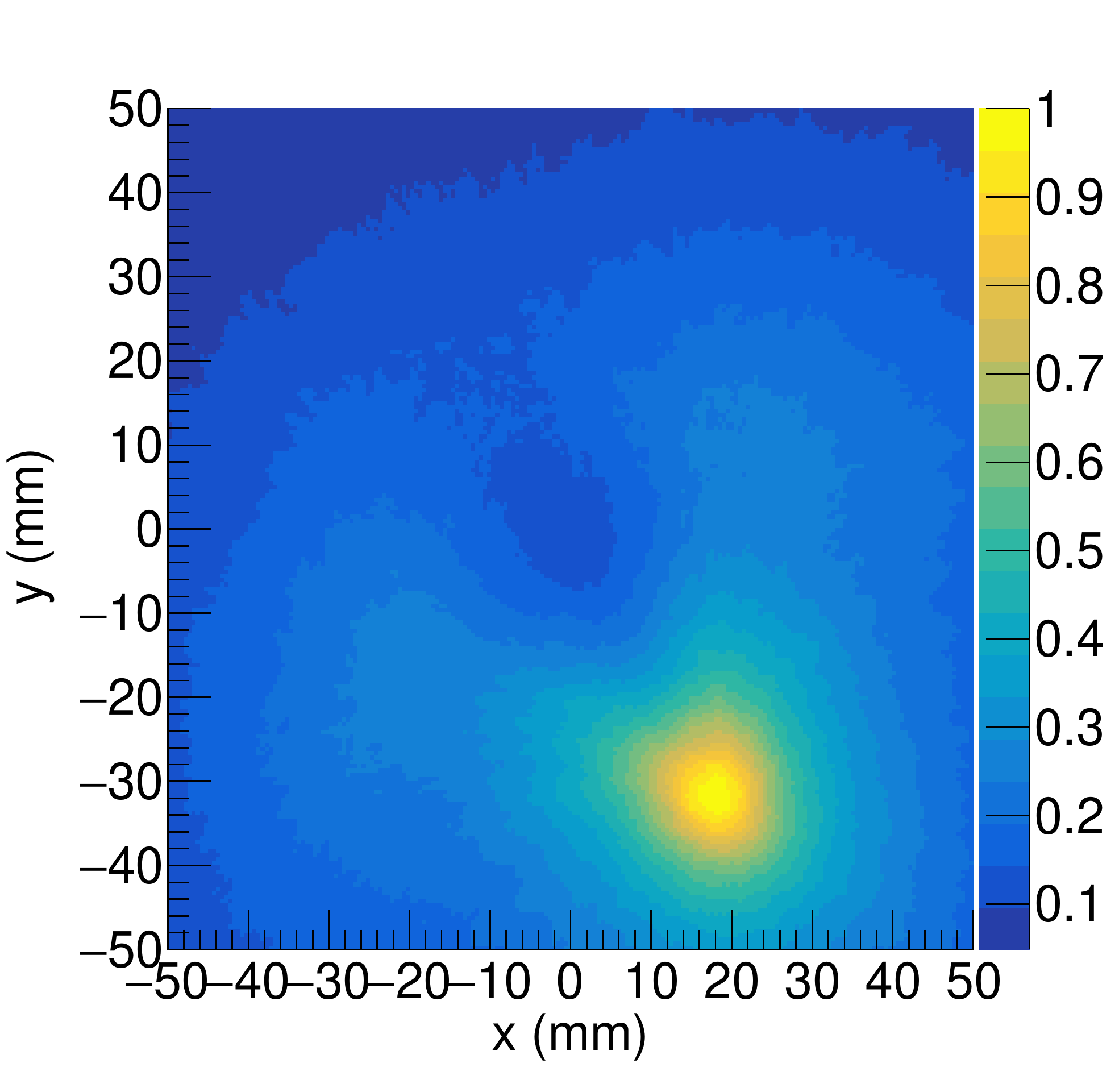}
	\includegraphics[width=0.3\textwidth]{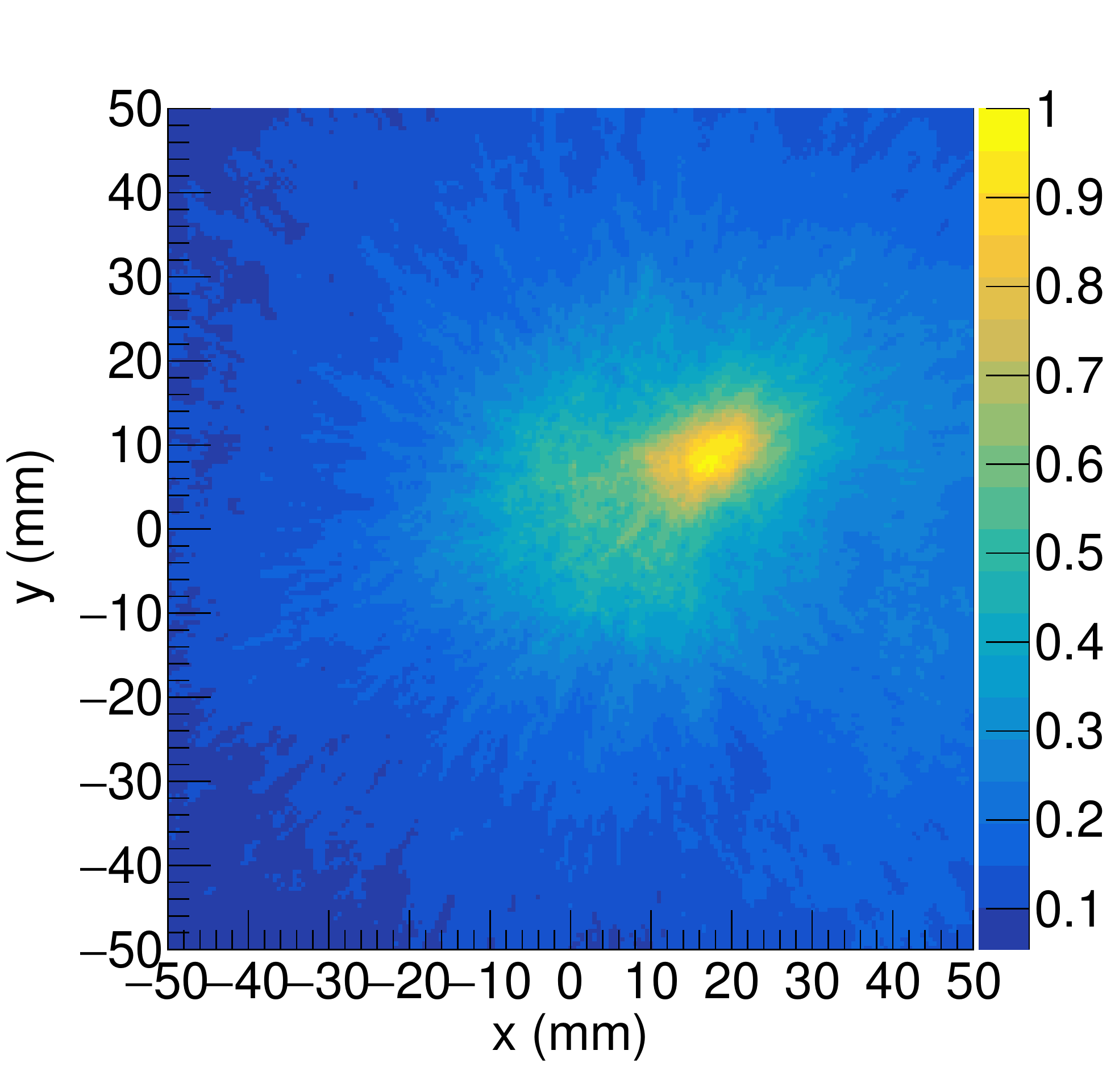}
	\includegraphics[width=0.3\textwidth]{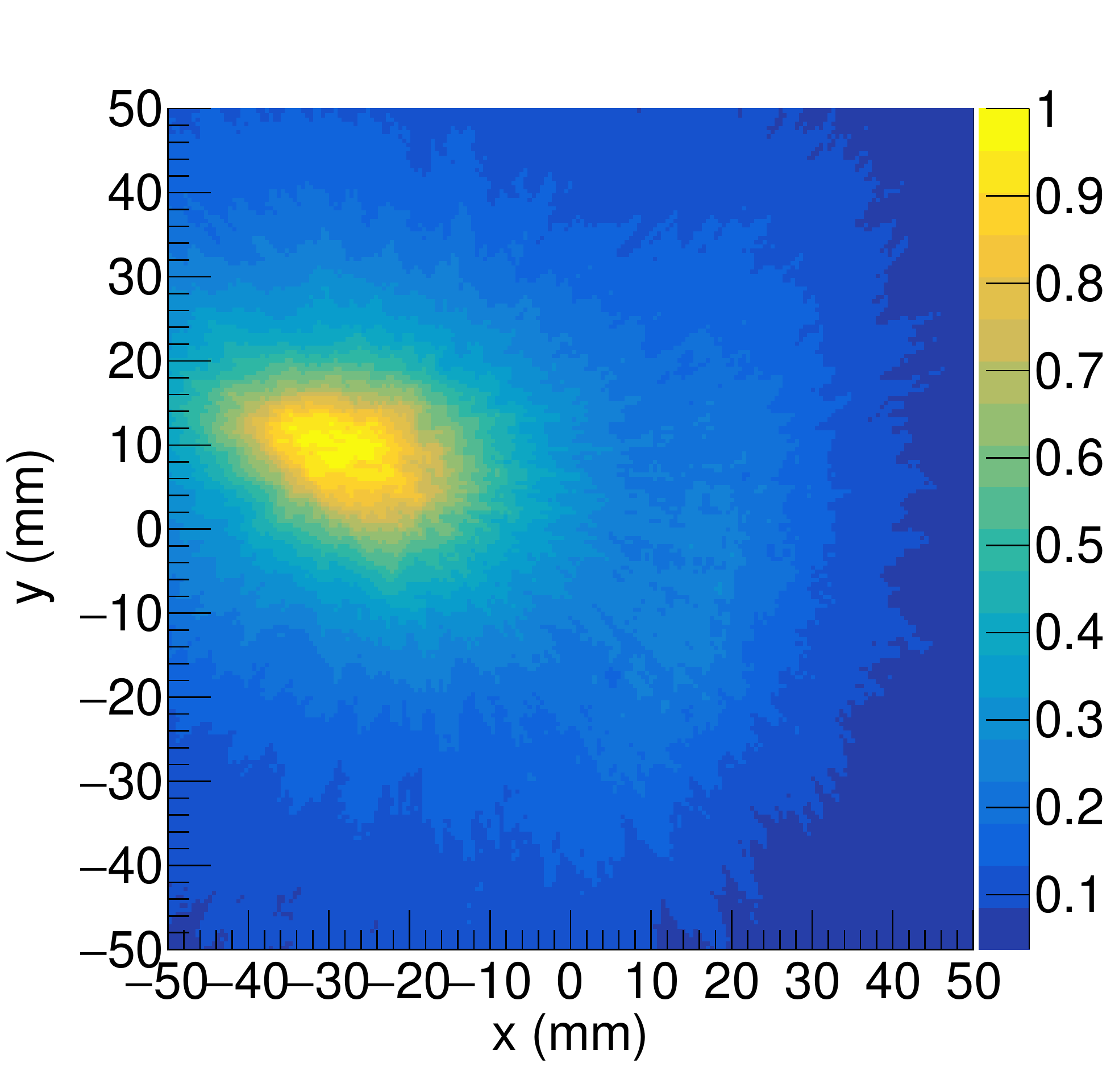}
	\includegraphics[width=0.3\textwidth]{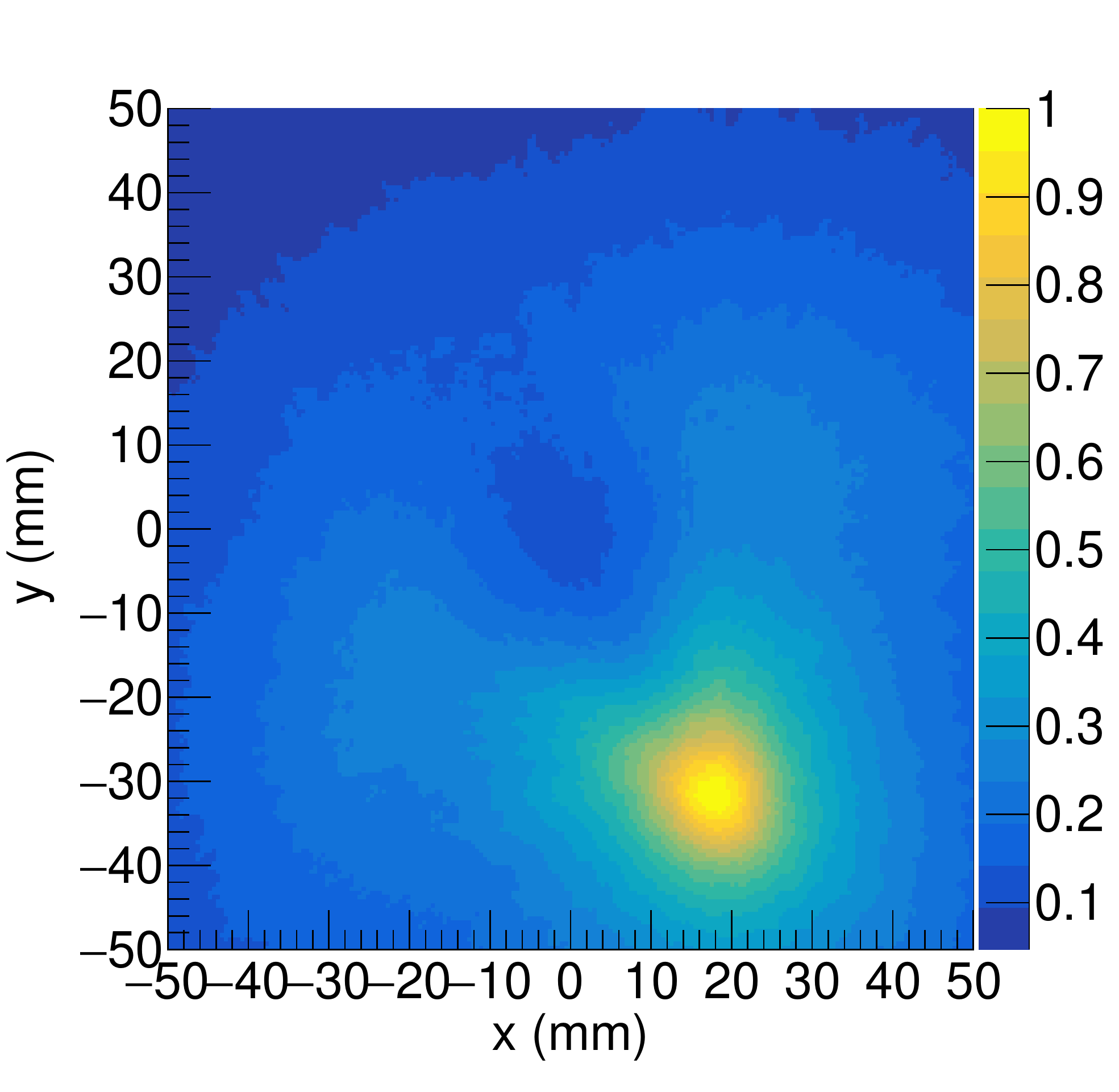}
	\includegraphics[width=0.3\textwidth]{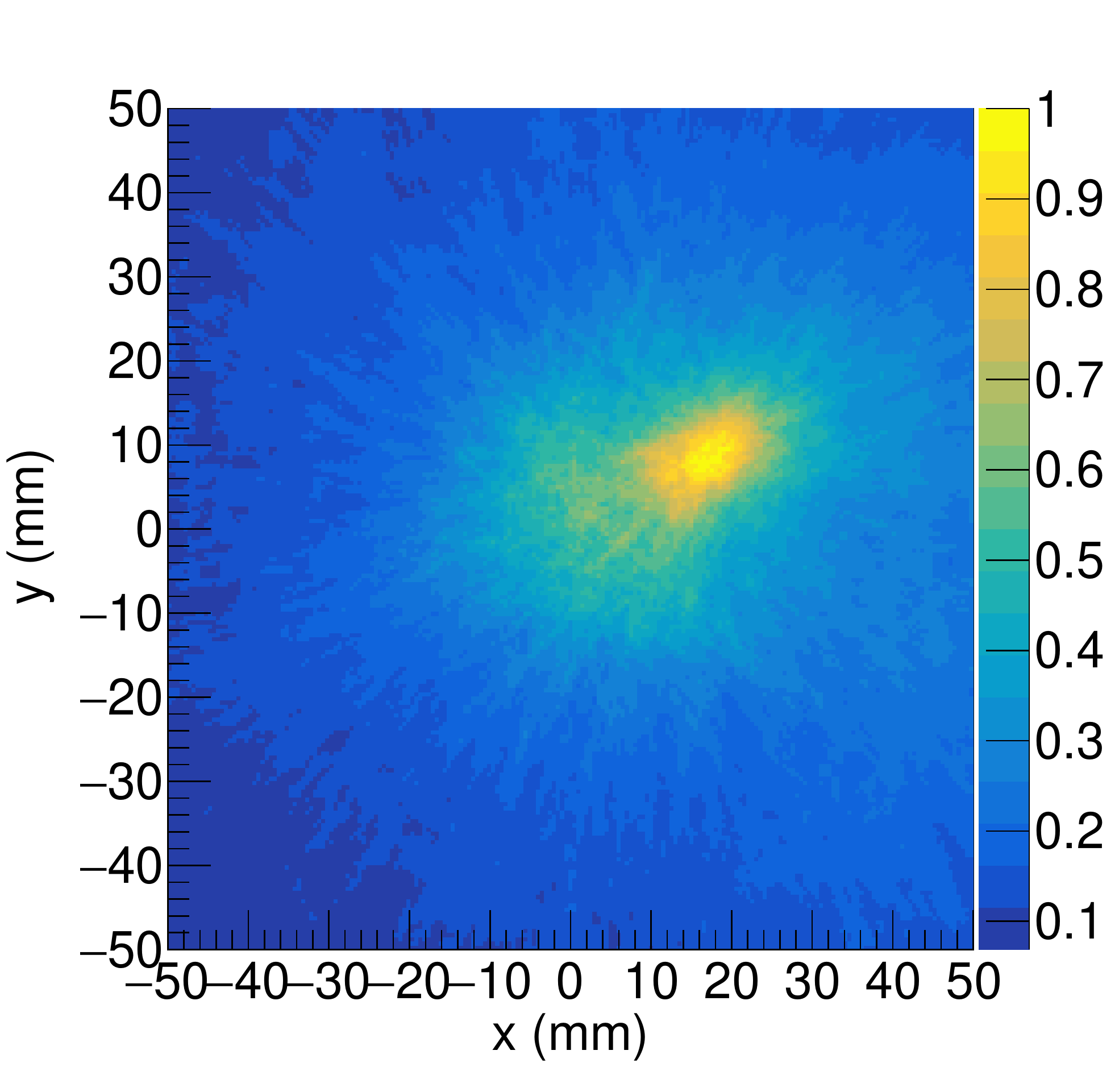}
	\includegraphics[width=0.3\textwidth]{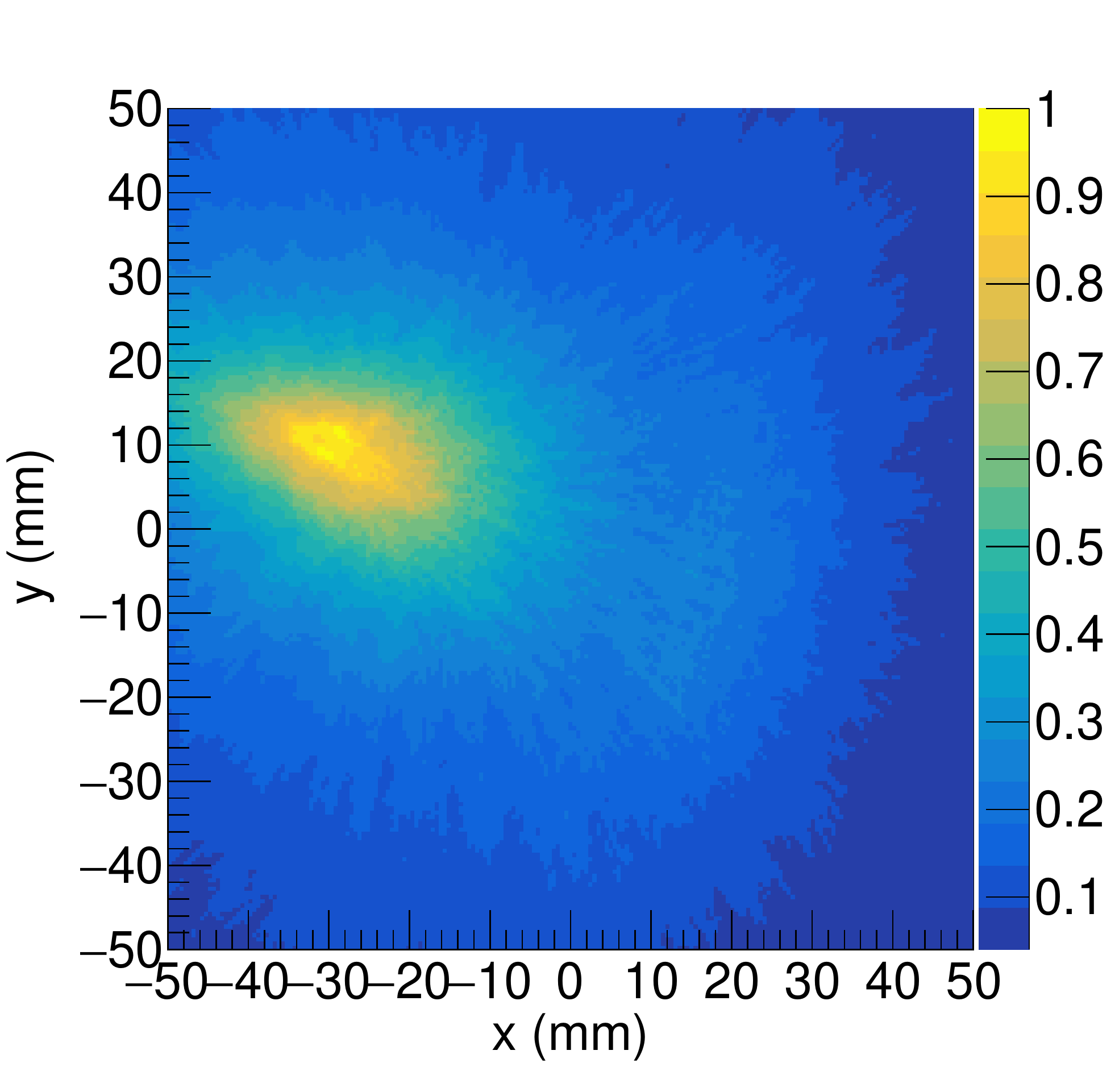}
	\caption{Images on the top: The interaction points are determined by the charge barycentre method in all the space clusters. Images on the bottom: The interaction points are reconstructed by the electron track algorithm in the space clusters with pixel numbers above 12. And the images shown from left to right are for energy windows of 320-400 keV, 620-700 keV, and 1100-1450 keV respectively.}
\label{fig:2DImage_MultiSource}
\end{figure}

To investigate the performance of the electron track algorithm for different gamma-ray sources with different energies and from different positions, the second measurement is performed with $\rm ^{137}Cs$, $\rm ^{133}Ba$, and $\rm ^{60}Co$ sources placed on the top of the sensor. The experimental setup is shown in \textcolor{black}{Fig.\ref{fig:ExperimentOne}}. These point-like sources are placed on a support and the vertical distance from the sources to the detector is 3.4 cm. Multiple energy windows, which are 320-400 keV, 620-700 keV, and 1100-1450 keV respectively, are applied on the measured data to select events to improve the image quality as shown in \textcolor{black}{Fig.\ref{fig:ExperimentOne}}. The selected time-coincidence clusters account for 0.26$\%$ of the total time-coincidence clusters, which is relatively low due to the compact size of the detector. The images of the multiple gamma-ray sources are reconstructed from a projection plane placed 3.4 cm above the detector. As shown in Fig.\ref{fig:2DImage_MultiSource}, the Compton images for each of the energy windows are reconstructed respectively. Fig.\ref{fig:2DImage_MultiSource} shows that the images of these gamma sources have been reconstructed correctly since these gamma-ray sources are located in the corresponding energy windows and positions.

\begin{figure}[!htb]
\centering
	\includegraphics[width=0.3\textwidth]{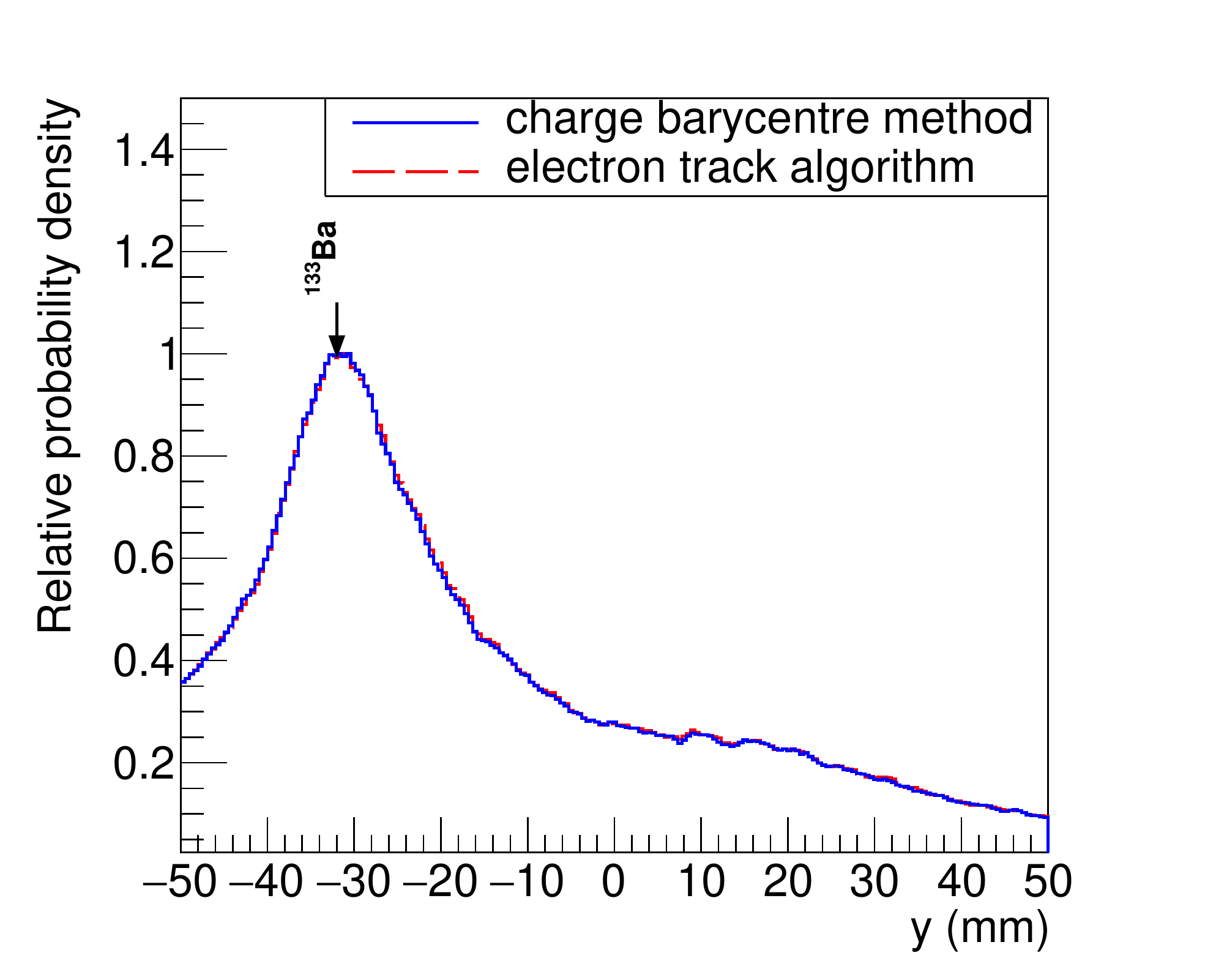}
	\includegraphics[width=0.3\textwidth]{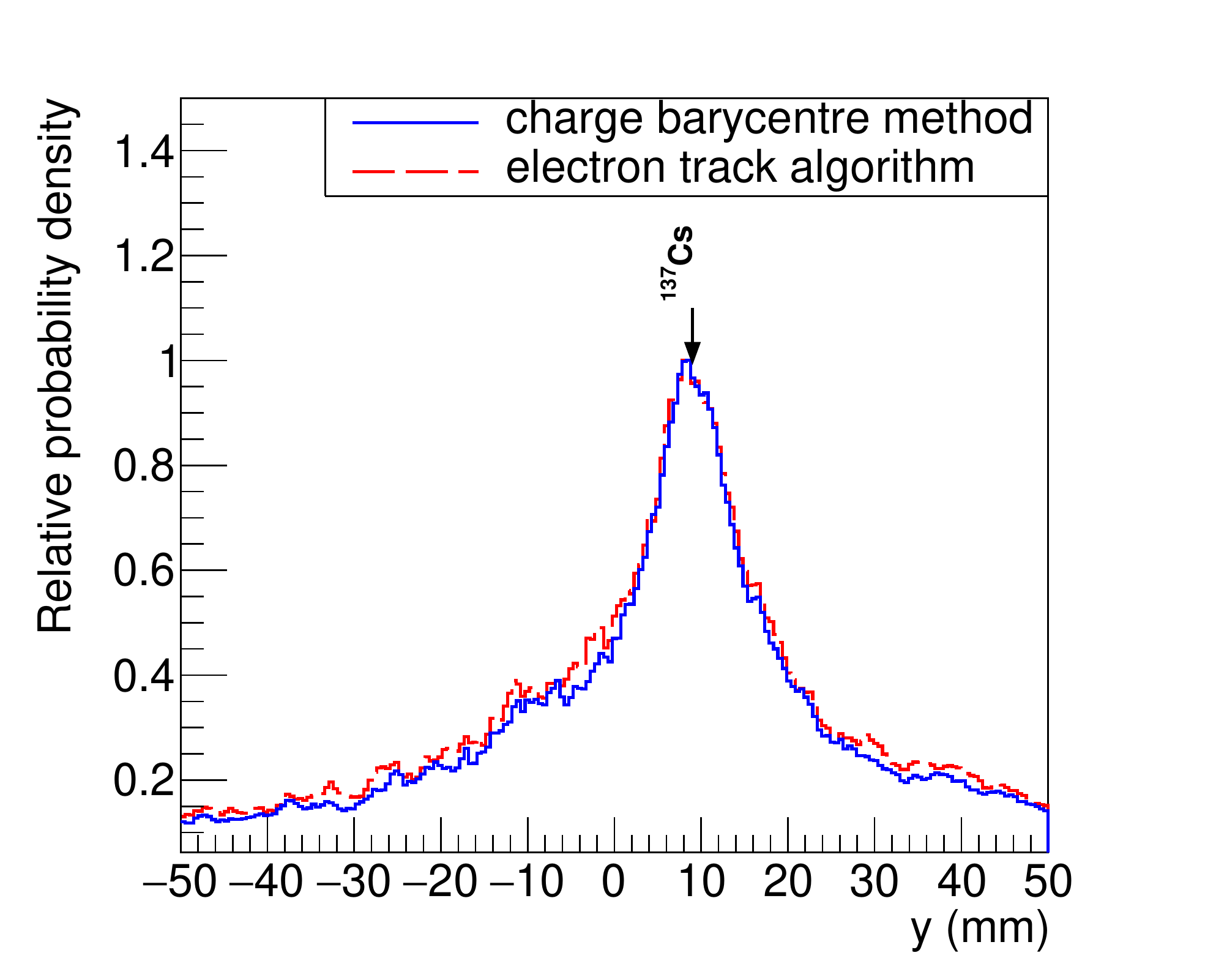}
	\includegraphics[width=0.3\textwidth]{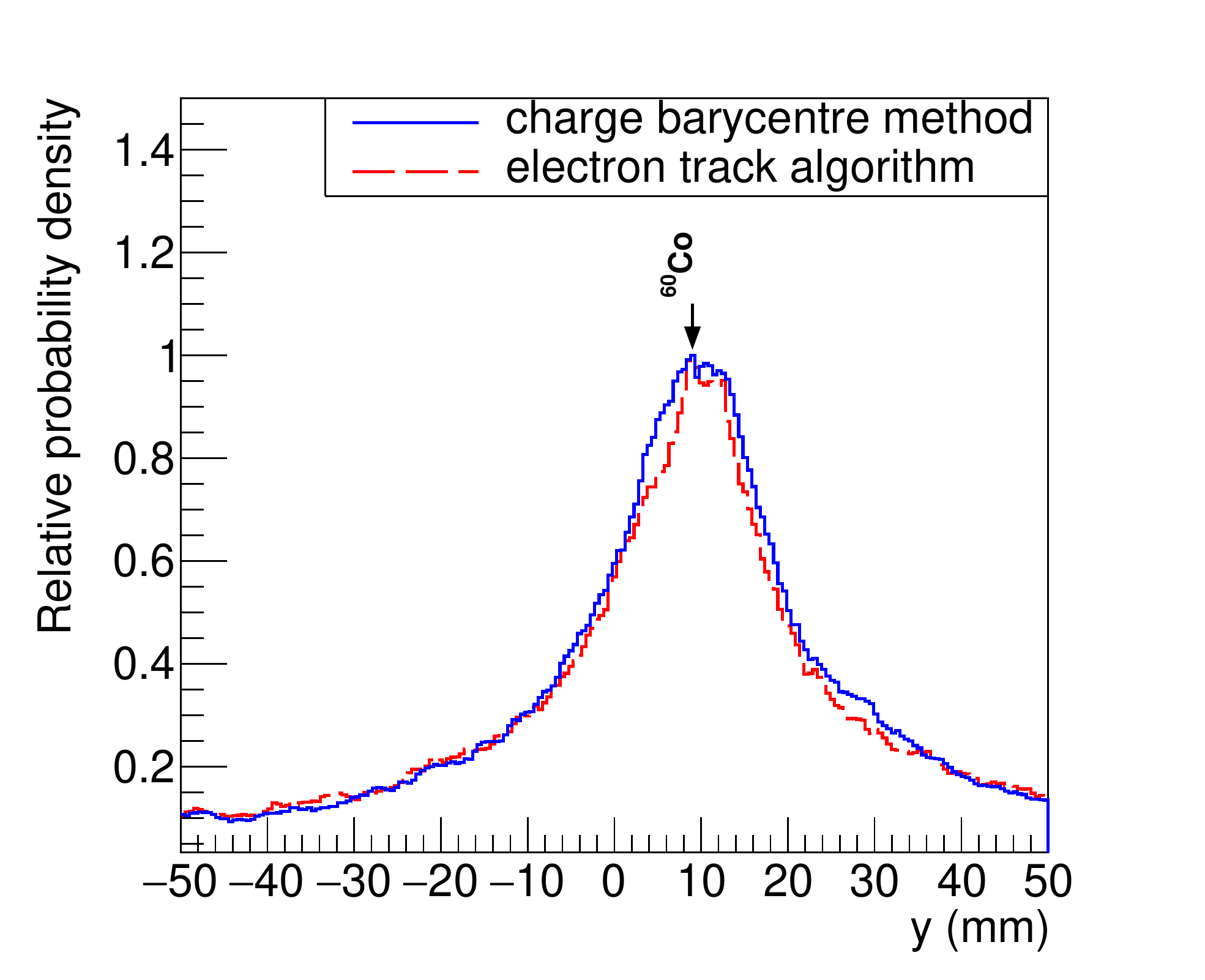}
	\caption{From left to right: Comparison of 1D probability density distributions along the y direction with x=18 mm and energy window of 320-400 keV, x=18 mm and energy window of 620-700 keV, and x=-30 mm and energy window of 1100-1450 keV, respectively.}
\label{fig:1DImage_MultiSource}
\end{figure}

By comparing the 1D probability density distributions of image reconstruction (see Fig.\ref{fig:1DImage_MultiSource}), an improvement of imaging resolution of $\rm ^{60}Co$ gamma-ray source can be seen and the FWHMs of the distributions are 22.6 mm for the charge barycentre method and 20.1 mm for the electron track algorithm. Compared with the results obtained in the first experiment, the electron track algorithm is effective for the $\rm ^{60}Co$ gamma-ray source placed at different positions. However, the electron track algorithm leads to a little degradation of the angular resolution of the $\rm ^{137}Cs$ gamma-ray source. We find the FWHM=16.1mm for the charge barycentre method and the FWHM=17.6 mm for the electron track algorithm, indicating that the interaction point of the Compton electrons generated by the 661.5 keV photons, which tend to have lower energies than those generated by the 1173 and 1332 keV photons, cannot be reconstructed precisely by the electron track algorithm. As for the $\rm ^{133}Ba$ gamma-ray source with an energy of 356 keV, the distributions are similar because there are very few electron tracks longer than 12 pixels generated by the 356 keV photons.

One of the disadvantages of this kind of compact Compton camera is the limited distance between the scattering point and absorption point, therefore, the precise reconstruction of the interaction point is of utter importance to determine the scatter direction. The results obtained in the Timepix3-based Compton camera indicate that the interaction points are reconstructed more precisely by the electron track algorithm, while the electrons are energetic and leave a long track in the detector, thus the angular resolution of the Compton camera can be improved. The Timepix3 we used has a sensor of CdTe and it has been shown that the electron track algorithm only has the effect of optimization for the $\rm ^{60}Co$ gamma-ray source with 1173 and 1332 keV gamma-ray lines. Since the silicon has a lower density than the CdTe, the continuous slowing down approximation (CSDA) range in silicon is approximate twice the CSDA range in CdTe for electrons with the same energy. And the silicon pixelated detector usually has a smaller pixel pitch. Therefore, in the silicon pixelated detector-based Compton camera, the electron track information could be more useful and an improvement of angular resolution in the hundreds of keV range is expected.

\section{Conclusion}
\label{sec:conclusion}
\textcolor{black}{The possibility of improving the image quality in conventional Compton camera} with the electron track reconstruction is demonstrated in this paper and experimentally verified with a pixelated semiconductor detector for the $\rm ^{60}Co$ gamma-ray source. Meanwhile, there are two new approaches applied for the Timepix3 data processing, the cosmic muon based charge induction time calibration and the graph theory based electron track reconstruction, which are both described in detail.



As illustrated in this paper, in imaging of a $\rm ^{60}Co$ gamma-ray source using the Timepix3-based single layer Compton camera, the angular resolution is significantly improved with the electron track algorithm, although the 55 $\rm \mu m$ pixel pitch is not sufficient for ETCC implementation. With the development of pixelated semiconductor detectors in near future, better energy resolution and spatial resolution can be expected. Conventional Compton camera with electron track algorithm in this paper, or even semiconductor-based ETCC, \textcolor{black}{would benefit from the better reconstruction of electron track and be implemented with the better angular resolution and hence the better sensitivity.}


\section{Acknowledgements}
This work was supported by the National Natural Science Foundation of China (Grant No. 11961141015, 11975214 and 12004353), Science Challenge Project (Grant No. TZ2018005) and National Key R\&D Program of China (Grant No. 2016YFA0401100).

\bibliography{Timepix3ComptonCamera}
\end{document}